\documentclass[11pt,a4paper]{amsart}
\usepackage[utf8]{inputenc}
\usepackage{amsmath}
\usepackage{array, amsfonts}
\usepackage{amsthm}
\usepackage{amssymb}
\usepackage{enumerate}
\usepackage{lineno}
\usepackage[bookmarks=true]{hyperref}
\usepackage{enumitem}
\usepackage{mathrsfs}
\usepackage{stackengine}
\usepackage{bookmark}
\usepackage{amsrefs}
\usepackage{color,soul}
\usepackage[draft]{todonotes}
\usepackage{fancyhdr}
\usepackage{MnSymbol}
\usepackage{algorithm2e}
\usepackage[noend]{algpseudocode}
\usepackage{mathtools}
\usepackage[foot]{amsaddr}

\definecolor{darkgreen}{rgb}{0,0.5,0}

\newcommand{\comment}[1]

\newcommand\myfunc[5]{%
	\begingroup
	\setlength\arraycolsep{0pt}
	#1\colon\begin{array}[t]{c >{{}}c<{{}} c}
		#2 & \to & #3 \\ #4 & \mapsto & #5 
	\end{array}%
	\endgroup}

\title{Algorithmic information\\ distortions and incompressibility\\ in\\ uniform multidimensional networks\footnote{In~\cite{Abrahao2018d}, preliminary results of this paper are presented as a Research Report no. 08/2018 at the National Laboratory for Scientific Computing (LNCC).}}

\author{Felipe S. Abrah\~{a}o}
\address[Felipe S. Abrah\~{a}o]{Centre for Logic, Epistemology and the History of Science (CLE), University of Campinas (UNICAMP), Brazil.}
\email{fabrahao@unicamp.br}
\address[Felipe S. Abrah\~{a}o, Klaus Wehmuth, Artur Ziviani]{National Laboratory for Scientific Computing (LNCC) -- Petropolis, RJ -- Brazil }
\address[Felipe S. Abrah\~{a}o, Hector Zenil]{Algorithmic Nature Group, Laboratoire de Recherche Scientifique (LABORES) for the Natural and Digital Sciences, Paris, France.}

\author[Klaus Wehmuth]{Klaus Wehmuth} 
\email{klaus@lncc.br}

\author{Hector Zenil}  
\address[Hector Zenil]{
Machine Learning Group, Department of Chemical Engineering and
Biotechnology, The University of Cambridge, U.K..\\
Oxford Immune Algorithmics, England, U.K.\\
Algorithmic Dynamics Lab, Unit of Computational Medicine, Department of Medicine Solna, Center for Molecular Medicine, Karolinska Institute, Stockholm, Sweden.\\
}
\email{hector.zenil@ki.se}

\author[Artur Ziviani]{Artur Ziviani}
\email{ziviani@lncc.br}

\fancypagestyle{plain}{
	\fancyhead{}
	\fancyhead[C]{ \textbf{Extended version of the paper:} }
}
\begin{document}

	\maketitle\thispagestyle{plain}
	
	
	\begin{abstract}\label{abstract}
		This article presents a theoretical investigation of generalized encoded forms of networks in a uniform multidimensional space. First, we study encoded networks with (finite) arbitrary node dimensions (or aspects), such as time instants or layers. In particular, we study these networks that are formalized in the form of multiaspect graphs. In the context of node-aligned non-uniform (or node-unaligned non-uniform and uniform) multidimensional spaces, previous results has shown that, unlike classical graphs, the algorithmic information of a multidimensional network is not in general dominated by the algorithmic information of the binary sequence that determines the presence or absence of edges. In the present work, first we demonstrate the existence of such algorithmic information distortions for node-aligned uniform multidimensional networks. Secondly, we show that there are particular cases of infinite nesting families of finite uniform multidimensional networks such that each member of these families is incompressible. From these results, we also recover the network topological properties and equivalences in irreducible information content of multidimensional networks in comparison to their isomorphic classical graph counterpart in the previous literature. 
		These results together establish a universal algorithmic approach and set limitations and conditions for irreducible information content analysis in comparing arbitrary networks with a large number of dimensions, such as multilayer networks.

	\end{abstract}
	
	
	

	\theoremstyle{definition}
	\newtheorem{notationundersubsection}{Notation}[subsection]
	\newtheorem{definitionafternotationundersubsection}[notationundersubsection]{Definition}
	\newtheorem{subdefinitionafternotationundersubsection}[notationundersubsection]{Definition}
	\newtheorem{definitionnotationafternotationundersubsection}[notationundersubsection]{Notation}
	\newtheorem{notationafternotationundersubsection}[notationundersubsection]{Notation}
	
	\theoremstyle{definition}
	\newtheorem{definitionundersection}{Definition}[section]

	\theoremstyle{remark}
	
	\theoremstyle{remark}
	\newtheorem{noteafternotationundersubsection}[notationundersubsection]{Note}

	\theoremstyle{theorem}
	\newtheorem{lemmaundersection}{Lemma}[section]
	\newtheorem{thmafterlemmaundersection}[lemmaundersection]{Theorem}
	\newtheorem{corollaryafterlemmaundersection}[lemmaundersection]{Corollary}

\section{Introduction}\label{sectionIntro}


Network complexity analysis and measures of information content of graphs by classical information-theoretic tools, e.g., Shannon entropy-like related measures, have been subjects of increasing importance in network modeling, network analysis or in data science in general \cite{Sole2004,Dehmer2011,Mowshowitz2012,Zenil2018a,Lewis2009}.
Not only for classical networks (i.e., monoplex networks), new models and methods are becoming crucial in the context of proper representations of multidimensional networks, such as dynamic networks and multilayer networks \cite{Kivela2014,Lambiotte2019,Michail2018,Wehmuth2016b}.
As a direct example from the source coding theorem \cite{Cover2005} in classical information theory, one has that every recursively structured data representation of a random graph $ G $ on $n$ vertices and edge probability $ p = 1/2 $ in the classical Erdős–Rényi model $ \mathcal{G}(n,p) $---from an independent and identically distributed stochastic process---is expected to be losslessly incompressible with probability arbitrarily close to one as $ \left| V(G) \right| = n \to \infty $ \cite{Li1997,Leung-Yan-Cheong1978}.
Nevertheless, applying statistic-informational measures to evaluate compressibility or computably irreducible information content of general data, such as strings or networks, may lead to deceiving measures in some cases~\cite{Cover2005,Zenil2018a,Zenil2017a,Li1997}.
For example, for some particular graphs displaying maximal degree-sequence entropy~\cite{Zenil2017a} or exhibiting a Borel-normal distribution of presence or absence of subsets of edges~\cite{Becher2002,Becher2015,Becher2013,Calude2016a}, the edge set $ E( G ) $ can be losslessly compressed into $ \mathbf{O}\left( \log(n) \right) $ bits \cite{Chaitin2004,Li1997,Downey2010,Calude2002}.

On the other hand, algorithmic information theory (AIT)~\cite{Li1997,Downey2019,Chaitin2004,Calude2002} (aka Kolmogorov complexity theory) has been given us a set of formal universal tools for studying data compression of individual (finite or infinite) objects~\cite{Downey2019,Barmpalias2019,Sow2003,Zenil2018,Delahaye2012,Li1997}, which are not necessarily constructed or defined by stochastic processes.
For example, this is the case of both a single algorithmically random infinite sequence or any particular computably generated object, such as computable strings, matrices, or graphs.
In addition, unlike data compression algorithm analysis, AIT enables worst-case compressibility and network complexity analyses that do not depend on the choice of programming language \cite{Zenil2018a,Morzy2017a}.
Moreover, formally grounded on computablity theory, information theory, and measure theory, such an algorithmic approach to the study of these objects (and not only graphs or networks but tensors in general) represents an important refinement of more traditional statistical approaches in applications to computer science and related subjects, such as network complexity (\cite{Morzy2017a,Zenil2018a}), machine learning (\cite{Zenil2019}), causality in complex systems (\cite{Zenil2019d,Zenil2018}), dynamical systems modeling (\cite{Rojas2008,Zenil2019d}), and thermodynamics of computatable systems (\cite{Zenil2019b,Tadaki2012,Baez2012}). 
For example, in the domain of the Principle Maximum Entropy \cite{Lesne2014} that helps build the underlying candidate ensemble under some constraints with the purposes of randomness deficiency estimation, the classical version is unable to distinguish recursive from non-recursive permutations~\cite{Zenil2019b}, potentially misleading an observer interested in unbiased features of an object (e.g.~\cite{Zenil2017a}) and comparisons to only non-recursive microstate configurations. 
Thus, being formally grounded on computablity theory, information theory, and measure theory, the algorithmic-informational approach has been contributing to new and more refined and robust investigations of the possible properties of mathematical objects, both statistical and algorithmic, 
in application to computer science.


In this way, the general scope of the present work is not only to study algorithmic complexity and algorithmic randomness of multidimensional networks, but also to study: the possible worst-case distortions from compressibility analyses in distinct multidimensional spaces; and the implications of incompressibility on networks' multidimensional topological properties.
We tackle the challenge by putting forward definitions, lemmas, theorems, and corollaries. 
Our results are based on previous applications of algorithmic information theory to string-based representations of classical graphs or networks \cite{Buhrman1999,Zenil2018a,Khoussainov2014,Abrahao2020cpublished,Abrahao2021publishednat}.
In addition, this article is based on multidimensional networks in the form of \emph{multiaspect graphs} (MAGs), as presented in \cite{Wehmuth2017,Wehmuth2016b}. 
These MAGs are formal representations of dyadic (or $2$-place) relational structures \cite{Hodges1993} between two arbitrary $n$-ary tuples. 
It has been shown that the MAG abstraction enables one to formally represent and computationally analyze networks with additional representational structures, e.g., dynamic networks \cite{Costa2015a,Wehmuth2015a,Brisaboa2018a} or dynamic multilayer networks \cite{Wehmuth2018,Wehmuth2016b,Wehmuth2018a}.
For having additional representational dimensions in which the nodes belong (or are ascribed to), e.g., time instants or layers, such networks are called multidimensional networks (or high-order networks) and have been shown to be of increasing overarching importance in complex systems science and network analysis~\cite{Kivela2014,Lambiotte2019,Michail2018}:
particularly, the study of dynamic  (i.e., time-varying) networks \cite{Rossetti2018b,Michail2015,Pan2011,Costa2015a}, multilayer networks \cite{Kivela2014,Boccaletti2014a,Domenico2013}, and dynamic multilayer networks \cite{Wehmuth2016b,Wehmuth2018a,Wehmuth2018}.

Contrary to previous work on algorithmic information and incompressibility for classical graphs \cite{Buhrman1999,Zenil2018a,Khoussainov2014},
previous work in \cite{Abrahao2020cpublished,Abrahao2021publishednat} demonstrated that isomorphisms between finite multidimensional networks and finite monoplex networks do not preserve algorithmic information in general.
This occurs because the irreducible information content of a multidimensional network may be highly dependent on the choice of its encoded isomorphic copy.
The underlying source of these distortions relies on the fact that the algorithmic information content of networks may be extremely sensitive to whether or not one is taking into account not only the total number of node dimensions but also the respective sizes of each node dimension, and~the ordering that they appear in the mathematical representation.
In other words, in the general case one needs additional irreducible information in order to compute the shape of the high-algorithmic-complexity multidimensional space. 

In particular,
when dealing with either \emph{uniform} or \emph{non-uniform} multidimensional spaces, it is shown in \cite{Abrahao2021publishednat} that \emph{node-unaligned} multidimensional networks can also display exponential algorithmic information distortions with respect to the algorithmic information content of their respective isomorphic monoplex networks.

It is also demonstrated that \emph{node-aligned} multidimensional networks with \emph{uniform} multidimensional spaces are limited to only displaying algorithmic information distortions that grow up to a logarithmic order of the number of extra node dimensions \cite{Abrahao2021publishednat}.
This differs from the node-aligned case studied in~\cite{Abrahao2020cpublished}, since worst-case distortions in the node-unaligned case are shown to grow at least exponentially with the number of extra node dimensions.
The algorithmic information content of node-aligned uniform multidimensional networks and the algorithmic information content of their isomorphic monoplex networks are tightly associated, except~maybe for a logarithmic factor of the number of extra node~dimensions.

However, only the upper bound is demonstrated in this previous works.
Thus, in the present article, first we construct an example and demonstrate the existence of a node-aligned uniform multidimensional network that display the above distortions. 
Secondly, we demonstrate that there are encodable infinite families of MAGs---and, consequentially, also of \emph{classical} (i.e., \emph{simple labeled}) graphs---that behave like families of algorithmically random strings (see Lemma~\ref{lemmaNestedfamilyofMAGs} and Theorem~\ref{thmNestedfamilyofK-randomMAGs}). 
We demonstrate that there is an algorithmic-informational cost within the isomorphisms between MAGs and their isomorphic graphs (see Section~\ref{subsectionMAGgraphsisomorphism}). 
By applying this algorithmic-information cost to previous work on algorithmically random classical graphs, one can retrieve some topological properties of such MAGs and families of MAGs, in particular, vertex degree, connectivity, diameter, and rigidity (see Section~\ref{sectionTopologicalproperties}).

To this end, as introduced in Section~\ref{sectionGeneralBackground}, we base our definitions and notations on previous work in algorithmic information theory \cite{Li1997,Downey2010,Calude2002,Chaitin2004}, MAGs and graph theory~\cite{Bollobas1998,Diestel2017,Brandes2005a,Wehmuth2017,Wehmuth2016b}, and in algorithmically random classical graphs \cite{Zenil2018a,Buhrman1999,Khoussainov2014}. 

In Section~\ref{sectionRecursivelylabeledMAGs}, we define \emph{encodable (i.e., recursively labeled)} MAGs and show how such mathematical objects are determined by the algorithmic information of arbitrarily chosen binary strings.
In fact, unlike classical graphs, the algorithmic information of a MAG and of the string that determines its (composite) edge set may be not so tightly associated regarding (plain or prefix) algorithmic complexity and mutual algorithmic information.
However, once we define \emph{recursively labeled infinite families} of MAGs in Section~\ref{subsectionRecursivelylabeledfamilyofMAGs} with a unique ordering for every possible (composite) edges, we will show that, in this case, both objects become of the same order in terms of algorithmic information.

In Section~\ref{sectionK-randommultiaspectgraph}, we introduce prefix algorithmic randomness (i.e., K-randomness) for finite MAGs and show that there are infinite families of finite MAGs (or classical graphs) in which every member is incompressible (i.e., K-random) regarding prefix algorithmic complexity (i.e., K-complexity). 
In addition, we show in Section~\ref{subsectionNestedMAGs} that there are recursively labeled infinite families of MAGs in which a member is a \emph{multiaspect subgraph} (subMAG) of the other. That is, such families are defined by an infinite sequence of MAGs such that the former is always a subMAG of the latter. Therefore, we will prove that one can obtain a \emph{recursively labeled infinite nesting family} of finite MAGs in which each member is as prefix algorithmically random as a prefix algorithmically random finite initial segment of infinite binary sequence.

In Section~\ref{sectionC-randomMAGs}, we relate these results on prefix algorithmic randomness with plain algorithmic randomness (i.e., C-randomness) of finite MAGs. 
Thus, as we show in Section~\ref{sectionTopologicalproperties}, this enables one to extend previous results on network topological properties in \cite{Li1997,Buhrman1999} to plain algorithmically random MAGs or prefix algorithmically random nesting families of MAGs.





\section{Background}\label{sectionGeneralBackground}

\subsection{Preliminary definitions and notations}\label{subsectionNotations}

\subsubsection{Graphs and multiaspect graphs }\label{subsubsectionGraphsandMAGs}

We directly base our notation regarding classical graphs on \cite{Diestel2017,Brandes2005a,Bollobas1998} and regarding multiaspect graphs on \cite{Wehmuth2016b,Wehmuth2017}.
In order to avoid ambiguities, minor differences in the notation from \cite{Wehmuth2016b,Wehmuth2017} will be introduced in this section.
First, as usual, let $ ( \, \cdot \, , \, \cdot \, ) $ denote an \emph{ordered pair}, which is defined by the cartesian product $ \times $  of two sets with cardinality $1$ each. 
Thus, the union of all these ordered pairs is the cartesian product of two sets $X$ and $Y$, where 
\[
x \in X \, \land \, y \in Y \iff (x,y) \in X \times Y
\text{ .}
\]
Now, let $ \{ \,  \cdot \, , \, \cdot \, \} $ denote a \emph{unordered pair}, which is a set with cardinality $2$. 
Then, one has the classical notion of graph:

%
%
%
%
%
%

\begin{definitionafternotationundersubsection}\label{BdefGraph}
	A \emph{labeled} (directed or undirected) \emph{graph}  $ G = ( V , E ) $ is defined by an ordered pair $ ( V , E ) $, where $ V= \{ 1 , \dots , n \} $ is the finite set of labeled vertices with $ n \in \mathbb{N} $ and $ E $ is the edge set such that 
	\[
	E \subseteq V \times V
	\]
\end{definitionafternotationundersubsection}	
%
	
Let $ V(G)$ denote the set of vertices of $G$.
Let $ E( G ) $ denote the edge set of $G$.
If a labeled graph $G$ does not contain \emph{self-loops}\footnote{ That is, there is no edge or arrow linking the same vertex to itself.}, i.e., for every $ x \in V $, 
\[
\left( x , x  \right) \notin E \text{ ,}
\]
\noindent then we say $G$ is a \emph{traditional graph}.
In addition, if the edges do not have orientation:

	\begin{subdefinitionafternotationundersubsection}\label{defClassicgraph}
		A labeled \emph{undirected} graph $ G = ( V , E )  $ without \emph{self-loops} is a labeled graph with a restriction $ \mathbb{E}_c $ in the edge set $E$ such that each edge is an \emph{unordered pair} with
		\[
		E \subseteq \mathbb{E}_c\left(  G \right) \coloneq \{ \{ x , y \} \mid x , y \in V \}  
		\] 
		\noindent where\footnote{ That is, the adjacency matrix of this graph is symmetric and the main diagonal is null. } there is $ Y \subseteq  V \times V $ such that
		\[
		\{ x , y \} \in E \subseteq \mathbb{E}_c\left(  G \right) \iff ( x , y ) \in Y \, \land \, ( y , x ) \in Y \, \land \, x \neq y
		\]
		\noindent We also refer to these graphs as \emph{classical} (or \emph{simple labeled}) graphs.
	\end{subdefinitionafternotationundersubsection}

For the present purposes of this article, and as classically found in the literature, all graphs $G$ will be classical graphs.

With regard to isomorphism and vertex permutation, one has the usual notion of rigidity in which one says a classical graph is \emph{rigid} if and only if its only automorphism is the identity automorphism.

%
%
%
%
	
	
%
	

%

One can also define graphs belonging to other graphs:

\begin{definitionafternotationundersubsection}\label{defSubgraphs}
	We say a graph $ G' $ is a \emph{subgraph} of a graph $ G $, denoted as $ G' \subseteq G $, \textit{iff}
	\[
	V\left( G' \right) \subseteq V\left( G \right) 
	\; \land \;
	E\left( G' \right) \subseteq E\left( G \right) 
	\]
	
\end{definitionafternotationundersubsection}
In particular, a graph $ G' $ is a \emph{vertex-induced subgraph} of $ G $ 	\textit{iff} 
\[
V\left( G' \right) \subseteq V\left( G \right) 
\] 
\noindent and, for every $ u , v \in V\left( G' \right)  $,
\[
( u , v ) \in E\left( G \right)  \, \implies \, ( u , v ) \in E\left( G' \right) 
\]
\noindent In addition, we denote this $ G' $ as $ G\left[ V\left( G' \right)  \right] $.

As established in \cite{Wehmuth2016b,Wehmuth2017}, we may generalize these notions of graph in order to represent dyadic (or $2$-place) relations between $n$-ary tuples:

\begin{definitionafternotationundersubsection}\label{defMAG}
	Let $ \mathscr{G}=(\mathscr{A},\mathscr{E}) $ be a multiaspect graph (MAG), where 
	\[
	\mathscr{E}( \mathscr{G} ) \subseteq \mathbb{E}(\mathscr{G})
	\] 
	is the set of existing composite edges of the MAG as in Notation~\ref{notationCompositeverticesandedges} and $\mathscr{A}$ is a finite list of sets, each of which is an \emph{aspect} denoted by $ \mathscr{A}( \mathscr{G} )[i] $, where $ 1 \leq i \leq p $.  
	
\end{definitionafternotationundersubsection} 
	
Each aspect $ \mathbf{ \sigma } \in \mathscr{A} $ is a finite set and the number of aspects $ p = | \mathscr{A} | $ is called the \emph{order} of $ \mathscr{G} $. By an immediate convention, we call a MAG with only one aspect as a \emph{first-order} MAG, a MAG with two aspects as a \emph{second-order} MAG and so on.  Each composite edge (or arrow) $ e \in \mathscr{E} $ may be denoted by an ordered $2p$-tuple $ ( a_1,\dots,a_p, b_1, \dots, b_p ) $, where $ a_i, b_i $ are elements of the $i$-th aspect with $ 1 \leq i \leq p = | \mathscr{A} | $.
Specifically, $ \mathscr{A}( \mathscr{G} ) $ denotes the class of aspects of $ \mathscr{G} $ and $ \mathscr{E}( \mathscr{G} ) $ denotes the \emph{composite edge set} of $ \mathscr{G} $.
We denote the $i$-th aspect of $ \mathscr{G} $ as $ \mathscr{A}( \mathscr{G} )[i] $. So, $ | \mathscr{A}( \mathscr{G} )[i] |$ denotes the number of elements in $ \mathscr{A}( \mathscr{G} )[i] $.
	
%
%
%
	
	In order to match the classical graph case, one may adopt the convention of calling the elements of the first aspect of a MAG as \emph{vertices}. 
	Therefore, one can denote the set $ \mathscr{A}( \mathscr{G} )[1] $ of elements of the first aspect of a MAG  $ \mathscr{G} $ by $ V( \mathscr{G} ) $. 
	Thus, a vertex should not be confused with a composite vertex:
	
	\begin{definitionnotationafternotationundersubsection}\label{notationCompositeverticesandedges}
		The set of all \emph{composite vertices} $ \mathbf{v} $ of $ \mathscr{G} $ is denoted by
		\[
		\mathbb{V}( \mathscr{G} ) = \bigtimes_{i=1}^{p} \mathscr{A}( \mathscr{G} )[i] 
		\]
		\noindent  and the set of all possible \emph{composite edges} $ e $ of $ \mathscr{G} $ is denoted by
		\[
		\mathbb{E}(\mathscr{G}) = \bigtimes_{n=1}^{2p}  \mathscr{A}(G)[ { (n-1) \pmod{p} } \, + 1 ]  
		= \mathbb{V}( \mathscr{G} ) \bigtimes \mathbb{V}( \mathscr{G} )
		\text{ ,}
		\]
		\noindent so that, for every ordered pair $ ( \mathbf{u} , \mathbf{v} ) $ with $ \mathbf{u} , \mathbf{v} \in \mathbb{V}( \mathscr{G} )  $, we have $ ( \mathbf{u} , \mathbf{v} ) = e \in \mathbb{E}( \mathscr{G} )  $. Also, for every $ e \in \mathbb{E}( \mathscr{G} )  $ we have $ ( \mathbf{u} , \mathbf{v} ) = e $ such that $ \mathbf{u} , \mathbf{v} \in \mathbb{V}( \mathscr{G} )  $.

%
		
	\end{definitionnotationafternotationundersubsection}
	
	The terms \emph{vertex} and \emph{node} may be employed interchangeably in this article. However, we choose to use the term \emph{node} preferentially within the context of networks, where nodes may realize operations, computations or would have some kind of agency, like in real-world networks. Thus, we choose to use the term \emph{vertex} preferentially in the mathematical context of graph theory. 
	
	\begin{definitionafternotationundersubsection}\label{defCompaniontuple}
		We denote the \emph{companion tuple} of a MAG $ \mathscr{G} $ as defined in \cite{Wehmuth2017} by $ \tau( \mathscr{G} ) $ where
		\[
		\tau( \mathscr{G} ) = \left( | \mathscr{A}( \mathscr{G} )[1] |, \dots , | \mathscr{A}( \mathscr{G} )[p] | \right)
		\]
		
%
		
	\end{definitionafternotationundersubsection}

As we will see in Notation~\ref{notationPairing}, let $ \left< \tau( \mathscr{G} )  \right> $ denote the string $ \left< | \mathscr{A}( \mathscr{G} )[1] |, \dots , | \mathscr{A}( \mathscr{G} )[p] | \right> $.
	
The diameter in the multidimensional case can also be defined in an analogous way to graphs as the maximum shortest path:
we define the composite diameter $ \mathrm{D}_\mathscr{E}( \mathscr{G} ) $ as the maximum value in the set of the minimum number of steps (through composite edges) in $ \mathscr{E}( \mathscr{G} ) $ necessary to reach a composite vertex $ \mathbf{v}$ from a composite vertex $ \mathbf{u} $, for any $  \mathbf{u} , \mathbf{v} \in \mathbb{V}( \mathscr{G} ) $.
See also \cite{Wehmuth2016b} for paths and distances in MAGs.

Analogously to traditional directed graphs in Definition~\ref{BdefGraph}, we have from \cite{Wehmuth2016b}: 

\begin{definitionafternotationundersubsection}\label{defTraditionalMAG}
	We define a \emph{directed} MAG $ \mathscr{G}_d = (\mathscr{A},\mathscr{E}) $ without \emph{self-loops} as a restriction $ \mathbb{E}_d $ in the set of all possible composite edges $ \mathbb{E} $ such that
	\[
	\mathscr{E}( \mathscr{G}_d )  \subseteq \mathbb{E}_d( \mathscr{G}_d ) 
	\coloneqq  
	\mathbb{E}(\mathscr{G}) \setminus \left\{ \left( \mathbf{u} , \mathbf{u} \right) \middle\vert \, \mathbf{u} \in \mathbb{V}( \mathscr{G}_d )  \right\}
	\]
	
	\noindent And we will have directly from this definition that
	\[
	\left| \mathbb{E}_d( \mathscr{G}_d )   \right| = { \left| \mathbb{V}( \mathscr{G}_d ) \right| }^2 - { \left| \mathbb{V}( \mathscr{G}_d ) \right| }
	\]
	
\end{definitionafternotationundersubsection}

We refer to these MAGs  $ \mathscr{G}_d $ in Definition~\ref{defTraditionalMAG} as \emph{traditional} MAGs.
Also note that a classical graph $G$, as in Definition \ref{defClassicgraph}, is a labeled first-order $ \mathscr{G}_d $ with $ \mathbb{V}( \mathscr{G}_d ) = \{ 1 , \dots , \left| \mathbb{V}( \mathscr{G}_d ) \right| \}$ and a symmetric adjacency matrix.
%
%
%
Thus, analogously to classical graphs in Definition~\ref{defClassicgraph}, we have:

\begin{definitionafternotationundersubsection}\label{defSimplifiedMAG}
	We define an \emph{undirected} MAG $ \mathscr{G}_c = (\mathscr{A},\mathscr{E}) $ without \emph{self-loops} as a restriction $ \mathbb{E}_c $ in the set of all composite edges $ \mathbb{E} $ such that
	\[
	\mathscr{E}( \mathscr{G}_c )  \subseteq \mathbb{E}_c( \mathscr{G}_c ) \coloneq   \{ \{ \mathbf{u} , \mathbf{v} \} \mid \mathbf{u},\mathbf{v} \in \mathbb{V}( \mathscr{G}_c )  \} 
	\]
	\noindent where there is $ Y \subseteq \mathbb{E}(\mathscr{G}_c)  $ such that
	\[
	\{ \mathbf{u} , \mathbf{v} \} \in \mathscr{E}( \mathscr{G}_c )  \iff 
	( \mathbf{u} , \mathbf{v} ) \in Y \, \land \, ( \mathbf{v} , \mathbf{u}) \in Y \land \mathbf{u} \neq \mathbf{v}
	\]
	\noindent And we have directly from this definition that
	\[
	\left| \mathbb{E}_c( \mathscr{G}_c )   \right| = \frac{ { \left| \mathbb{V}( \mathscr{G}_c ) \right| }^2 - { \left| \mathbb{V}( \mathscr{G}_c ) \right| } }{ 2 }
	\text{ .}
	\]
	We refer to these MAGs  $ \mathscr{G}_c $ in Definition~\ref{defSimplifiedMAG} as \emph{simple} MAGs.
	
%
%
%
\end{definitionafternotationundersubsection}

Thus, note that a classical graph $G$, as in Definition \ref{defClassicgraph}, is a labeled first-order simple MAG $ \mathscr{G}_c $ with $ \mathbb{V}( \mathscr{G}_c ) = \{ 1 , \dots , \left| \mathbb{V}( \mathscr{G}_c ) \right| \}$.

As in \cite{Wehmuth2016b}, one can define a MAG-graph isomorphism analogously to the classical notion of graph isomorphism:

\begin{definitionafternotationundersubsection}\label{defMAGandgraphsisomorphisms}
	We say a traditional MAG $ \mathscr{G}_d $  is isomorphic to a traditional directed graph $ G $ when there is a bijective function $ f : \mathbb{V}( \mathscr{G}_d ) \to V( G ) $ such that 
	\[ e \in \mathscr{E}( \mathscr{G}_d ) \iff ( f( \pi_o( e ) ) , f( \pi_d( e ) ) ) \in E( G ) \text{ ,}\]
	where $ \pi_o $ is a function that returns the origin composite vertex of a composite edge and $ \pi_d $ is a function that returns the destination composite vertex of a composite edge.
\end{definitionafternotationundersubsection}

Note that Definition~\ref{defMAGandgraphsisomorphisms} applies analogously to isomorphism between simple MAGs and their respective classical graphs.

In order to avoid ambiguities with the classical isomorphism in graphs from vertex labels transformations, we call such an isomorphism between a MAG and graph from \cite{Wehmuth2016b} a \emph{MAG-graph isomorphism}, the usual isomorphism between graphs (\cite{Diestel2017,Bollobas1998}) as \emph{graph isomorphism}, and the isomorphism between two MAGs $ \mathscr{G} $ and $ \mathscr{G}' $ (i.e., $ ( \mathbf{ u } , \mathbf{ v } ) \in \mathscr{E}( \mathscr{G} ) \iff ( f( \mathbf{ u } ) , f( \mathbf{ v } ) ) \in \mathscr{E}( \mathscr{G}' ) $ ) as \emph{MAG isomorphism}. 
With regard to graph isomorphism or vertex permutation, one has the usual notion of rigidity in which one says a classical graph is \emph{rigid} if and only if its only graph automorphism is the identity graph automorphism and the same analogous notion of rigidity applies with regard to MAG automorphisms (see e.g. Corollaries~\ref{corC-randomMAGs} and~\ref{corK-randomnestedfamilyproperties}).

\subsubsection{Turing machines and algorithmic information theory}\label{subsubsectionTMandAIT}

In this section, we recover notations and definitions from the literature regarding algorithmic information theory (aka Kolmogorov complexity theory or Solomonoff-Kolmogorov-Chaitin complexity theory).
For an introduction to these concepts and notation, see \cite{Li1997,Downey2010,Calude2002,Chaitin2004}.

First, regarding some basic notation, 
let $ \{ 0 , 1 \}^* $ be the set of all binary strings (i.e., of all finite binary sequences).
Let $ l(x) $ denote the length of a string $ x \in  \{ 0 , 1 \}^* $.\footnote{ In \cite{Downey2010}, $ l(x) $ is denoted by $ |x| $. However, we choose here to leave $ |x| $ to denote the cardinality (or size) of $x$. }
Let $ (x)_2 $ denote the binary representation of the number $ x \in \mathbb{N} $. 
In addition, let $ (x)_{L} $ denote the representation of the number $ x \in \mathbb{N} $ in language $ L $.
Let $ x \upharpoonright_{n} $ denote the ordered sequence of the first $n$ bits of the fractional part in the binary representation of $ x \in \mathbb{R} $. That is, $ x \upharpoonright_{n} = x_1 x_2 \dots x_n \equiv (x_1, x_2, \dots, x_n ) $, where $ { ( x ) }_2 = y.x_1 x_2 \dots x_n x_{n+1} \dots $ with $ y \in \{ 0 , 1 \}^* $ and $  x_1 , x_2 , \dots , x_n \, \in \{ 0 , 1 \} $.
Let $\lg(x)$ denote the binary logarithm $\log_{2}(x)$.
%

%
%
%
%
%
%

%

\begin{notationundersubsection}\label{BdefFunctionU}
	Let $ \mathbf{U}(x) $ denote the output of a universal Turing machine $\mathbf{U}$ when $x$ is given as input in its tape. Thus, $ \mathbf{U}(x) $ denotes a \emph{partial recursive} function
	\[
	\myfunc{ \varphi_{\mathbf{U}} }{ L }{ L }{ x }{ y = \varphi_{\mathbf{U}}(x) } \text{ ,}
	\]  
	\noindent where $L$ is a language.

\end{notationundersubsection}

In particular, $ \varphi_{\mathbf{U}}(x) $ is a \emph{universal} partial recursive function \cite{Rogers1987,Li1997}. 
Note that, if $x$ is a non-halting program on $\mathbf{U}$, then this function $\mathbf{U}(x)$ is undefined for $x$.
Wherever $ \, n \in \mathbb{N} $ or $ n \in \{ 0 , 1 \}^* $ appears in the domain or in the codomain of a partial (or total) recursive function
\[
\myfunc{ \varphi_{ \mathcal{U} } }{ L }{ L }{ x }{ y = \varphi_{ \mathcal{U} }(x) } \text{ ,}
\]
\noindent where $ \mathcal{U} $ is a Turing machine, running on language $L$, it actually denotes $ \left( n \right)_{ L } $.
%
Let $ \mathbf{L_U} $ denote a binary\footnote{ Possibly, also containing delimiter symbols.} universal programming language for a universal Turing machine $\mathbf{U}$.
Let $ \mathbf{L'_U} $ denote a binary \emph{prefix-free} (or \emph{self-delimiting}) universal programming language for a prefix universal Turing machine $\mathbf{U}$.
Note that, although the same letter $ \mathbf{U} $ is used in the prefix-free case and in the plain universal case, the two universal Turing machines may be different, since, for $ \mathbf{L_U} $, the Turing machine does not need to be prefix-free. Thus, every time the domain of function $ \mathbf{U}(x) $ is in $ \mathbf{L_U} $, $ \mathbf{U} $ denotes an arbitrary universal Turing machine. Analogously, every time the domain of function $ \mathbf{U}(x) $ is in $ \mathbf{L'_U} $, $ \mathbf{U} $ denotes a \emph{prefix} universal Turing machine. If $ \mathbf{L'_U} $ or $ \mathbf{L_U} $ are not being specified, one may assume an arbitrary universal Turing machine.

As in \cite{Li1997,Downey2010}:

\begin{notationundersubsection}\label{notationPairing}
	Let $ \left< \, \cdot \, , \, \cdot \, \right> $ denote an arbitrary recursive bijective pairing function. 
	
\end{notationundersubsection}

This notation can be recursively extended to $ \left<   \, \cdot \, ,  \left< \, \cdot \, , \, \cdot \, \right> \right> $ and, then, to an ordered $3$-tuple $ \left< \, \cdot \, , \, \cdot \,  \, , \, \cdot \,\right> $. 
Thus, this iteration can be recursively applied with the purpose of defining self-delimited finite ordered $n$-tuples $ \left< \cdot \, , \, \dots \, , \, \cdot   \right> $, so that there is only one natural number univocally representing a particular $ n $-tuple, where $ n \in \mathbb{N} $.
In addition, one may assume a pairing function---as, for example, the recursively functionalizable concatenation ``$\circ$'' in \cite{Abrahao2016}---in which there is $ p \in \{  0 , 1 \}^{*} $ such that, for every $ n $-tuple $ \left< a_1 \, , \, \dots \, , \, a_n   \right> $, where $ n \in \mathbb{N} $, 
\[ \mathbf{U}\left( \left<  \left< a_1 \, , \, \dots \, , \, a_n   \right>  , p \right> \right) = n  \text{ .}\]

Now, we can restate the fundamental definitions of algorithmic complexity theory:

\begin{definitionafternotationundersubsection}\label{BdefC}
	The (unconditional) \emph{plain} \emph{algorithmic complexity} (also known as C-complexity, plain Kolmogorov complexity, plain program-size complexity, or Solomonoff-Kolmogorov-Chaitin complexity for universal Turing machines) of a string $ w $, denoted by $ C(w) $, is the length of the shortest program $w^* \in \mathbf{L_U} $ such that $ \mathbf{U}(w^*) = w $.\footnote{ $ w^* $ denotes the lexicographically first $ \mathrm{p} \in \mathbf{L_U} $ such that $ l(\mathrm{p}) $ is minimum and $ \mathbf{U}(p) = w $.} The \emph{conditional} plain algorithmic complexity of a string $ y $ given a string $ x $, denoted by $ C( y \, | x ) $, is the length of the shortest program $w \in \mathbf{L_U} $ such that $ \mathbf{U}( \left< x , w \right> ) = y $. Note that $ C( y ) = C( y \, | \epsilon ) $, where $ \epsilon $ is the empty string. We also have the \emph{joint} plain algorithmic complexity of strings $x$ and $y$ denoted by $ C( x , y ) \coloneq C( \left< x , y \right> ) $ and the \emph{C-complexity of information} in $x$ about $y$ denoted by $ I_C( x : y ) \coloneq C(y) - C( y \, | x ) $.
\end{definitionafternotationundersubsection}

This way, one can promptly apply pairing functions with the purpose of univocally encoding a MAG into a sequence: 	
	
	\begin{notationafternotationundersubsection}\label{BdefPlaincomplexityofedgesets}
		Let $ \left( e_1 , \dots , e_{ \left| \mathbb{E}( \mathscr{G} )  \right| } \right) $ be a previously fixed ordering (or indexing) of the set $ \mathbb{E}( \mathscr{G} ) $.
		For an (composite) edge set $ \mathscr{E}( \mathscr{G} ) $, let $ C( \mathscr{E}( \mathscr{G} ) ) \coloneq C( \left< \mathscr{E}( \mathscr{G} ) \right> ) $, where $\left< \mathscr{E}( \mathscr{G} ) \right>$ denotes the (composite) edge set string
		\[
		\left<  \left< e_1, z_1 \right>, \dots , \left< e_n , z_n \right>  \right> 
		\]
		\noindent such that 
		\[
		z_i = 1  \iff e_i \in \mathscr{E}( \mathscr{G} )
		\text{ ,}
		\]
		\noindent where  $ z_i \in \{ 0 , 1 \} $ with $ 1 \leq i \leq n= | \mathbb{E}( \mathscr{G} ) | $. 
		Thus, in the simple MAG case (as in Definition~\ref{defSimplifiedMAG}) with the ordering fixed (as in Definition~\ref{BdefCharacteristicstringofasimpleMAG}), we will have that $\left< \mathscr{E}( \mathscr{G}_c ) \right>$ denotes the (composite) edge set string
		\[
		\left<  \left< e_1, z_1 \right>, \dots , \left< e_n , z_n \right>  \right> 
		\]
		\noindent such that 
		\[
		z_i = 1  \iff e_i \in \mathscr{E}( \mathscr{G}_c )
		\text{ ,}
		\]
		\noindent where  $ z_i \in \{ 0 , 1 \} $ with $ 1 \leq i \leq n= | \mathbb{E}_c( \mathscr{G}_c ) | $.
		The same applies analogously to the conditional, joint, and C-complexity of information cases from Definition~\ref{BdefC}.
	\end{notationafternotationundersubsection}

And, for prefix-free or self-delimiting languages, we have:

\begin{definitionafternotationundersubsection}\label{BdefK}
	The (unconditional) \emph{prefix} \emph{algorithmic complexity} (also known as K-complexity, prefix Kolmogorov complexity, prefix-free program-size complexity, or Solomonoff-Kolmogorov-Chaitin complexity for prefix universal Turing machines) of a binary string $ w $, denoted by $ K(w) $, is the length of the shortest program $w^* \in \mathbf{L'_U} $ such that $ \mathbf{U}(w^*) = w $.\footnote{ $ w^* $ denotes the lexicographically first $ \mathrm{p} \in \mathbf{L'_U} $ such that $ l(\mathrm{p}) $ is minimum and $ \mathbf{U}(p) = w $.} 
	The \emph{conditional} prefix algorithmic complexity of a binary string $ y $ given a binary string $ x $, denoted by $ K( y \, | x ) $, is the length of the shortest program $w \in \mathbf{L'_U} $ such that $ \mathbf{U}( \left< x , w \right> ) = y $. Note that $ K( y ) = K( y \, | \epsilon ) $, where $ \epsilon $ is the empty string. 
	We have the \emph{joint} prefix algorithmic complexity of strings $x$ and $y$ denoted by $ K( x , y ) \coloneqq K( \left< x , y \right> ) $, the \emph{K-complexity of information} in $x$ about $y$ denoted by $ I_K( x : y ) \coloneq K(y) - K( y \, | x ) $, and the \emph{mutual algorithmic information} of the two strings $x$ and $y$ denoted by $ I_A( x \, ; y ) \coloneq K(y) - K( y \, | x^* ) $.

	\begin{notationafternotationundersubsection}\label{BdefPrefixcomplexityofedgesets}
		Analogously to Notation~\ref{BdefPlaincomplexityofedgesets}, for an (composite) edge set $ \mathscr{E}( \mathscr{G} ) $, let $ K( \mathscr{E}( \mathscr{G} ) ) \coloneq K( \left< \mathscr{E}( \mathscr{G} ) \right> ) $ denote 
		\[
		K( \left<  \left< e_1, z_1 \right>, \dots , \left< e_n , z_n \right>  \right> )
		\]
		\noindent such that 
		\[
		z_i = 1  \iff e_i \in \mathscr{E}( \mathscr{G} )
		\text{ ,}
		\]
		\noindent where  $ z_i \in \{ 0 , 1 \} $ with $ 1 \leq i \leq n= | \mathbb{E}( \mathscr{G} ) | $. The same applies analogously to the simple MAG case (as in Notation~\ref{BdefPlaincomplexityofedgesets}) and the conditional, joint, K-complexity of information, and mutual cases from Definition~\ref{BdefK}.
	\end{notationafternotationundersubsection}
\end{definitionafternotationundersubsection}

Then, we turn our attentions to algorithmic randomness:

\begin{definitionafternotationundersubsection}\label{defWeakK-randomnessofstrings}
	We say a string $ x \in \{ 0 , 1 \}^* $ is \emph{weakly K-random} (\emph{K-incompressible up to a constant}, \emph{$c$-K-incompressible}, \emph{prefix algorithmically random up to a constant} or \emph{prefix-free incompressible up to a constant}) if and only if, for a fixed constant $ d \in \mathbb{N} $,
	\[
	K( x ) \geq l( x ) - d
	\text{ .}
	\]
	
\end{definitionafternotationundersubsection}

With respect to weak asymptotic dominance of function $f$ by a function $g$, 
we employ the usual $f(x)=\mathbf{O}( g(x) )$ for the big \textbf{O} notation when $f$ is asymptotically upper bounded by $g$; 
and with respect to strong asymptotic dominance by a function $g$, we employ the usual $f(x)=\mathbf{o}( g(x) )$ when $g$ dominates $f$. 
As one of the conceptual pillars of algorithmic information theory, algorithmic randomness (or \emph{Martin-Löf randomness}) also allows one to study the incompressibility of infinite objects, such as real numbers with an infinite fractional part in their binary representation:

\begin{definitionafternotationundersubsection}\label{defK-randomnessofreals}
	We say a real number\footnote{ Or infinite binary sequence.} $ x \in \left[ 0 , 1 \right] \subset \mathbb{R} $ is \emph{1-random} (\emph{K-random}, or \emph{prefix algorithmically random}) if and only if it satisfies
	\[
	K( x \upharpoonright_n ) \geq n - \mathbf{O}(1)
	\text{ ,}
	\]
	\noindent where $ n \in \mathbb{N} $ is arbitrary.
	
%
%
%
\end{definitionafternotationundersubsection}

In order to avoid ambiguities between plain and prefix algorithmic complexity and ambiguities in relation to randomness deficiencies, we choose to say that a $1$-random real number (i.e., an algorithmically random real number with respect to prefix algorithmic complexity) in Definition~\ref{defK-randomnessofreals} is \emph{$ \mathbf{O}(1) $-K-random}.
Thus, a real number $ x \in \left[ 0 , 1 \right] \subset \mathbb{R} $ is $ \mathbf{O}(1) $-K-random 	\textit{iff} it is weakly K-random by a maximum $ d \in \mathbb{N} $ for every initial segment $ x \upharpoonright_n $ \cite{Downey2010}.

\subsubsection{Algorithmically random graphs}\label{subsubsectionARGdefinitions}

From \cite{Li1997,Buhrman1999}, we restate the definition of a labeled finite graph that has a randomness deficiency at most $ \delta(n) $ under a string-based representation:

\begin{definitionafternotationundersubsection}\label{defC-randomgraph}
	
	A classical graph $ G $  with $ | V(G) | = n $ is $ \delta(n) $-random if and only if it satisfies
	\[
	C( E(G) \, | n ) \geq \binom{n}{2} - \delta(n)
	\text{ ,}
	\]
	\noindent where 
	\[ \myfunc{ \delta }{ \mathbb{N} }{ \mathbb{N} }{ n  }{ \delta(n) } \] 
	\noindent is a randomness deficiency function.
	
%
\end{definitionafternotationundersubsection}
In order to avoid ambiguities between plain and prefix algorithmic complexity, we choose to say that a $\delta(n)$-random graph $G$ in Definition~\ref{defC-randomgraph} is \emph{$ \delta(n) $-C-random}.

\subsection{Previous results}\label{subsectionBackground}

\subsubsection{Multiaspect graphs}\label{subsubsectionMultiaspect-graphs}

This section restates some previous results in \cite{Wehmuth2017,Wehmuth2016b}. First, it has been shown that a MAG is basically equivalent to a traditional directed graph, except for an isomorphism between relabeling vertices as composite vertices, and vice-versa \cite{Wehmuth2016b}:

\begin{thmafterlemmaundersection}\label{thmMAGisomorphism}
	For every traditional MAG $ \mathscr{G}_d $ of order $p>0$, where all aspects are non-empty sets, there is a unique (up to a graph isomorphism) traditional directed graph $ G_{ \mathscr{G}_d } = \left( V , E \right) $ with $ | V( G ) | =  \prod\limits_{ n = 1 }^{ p } | \mathscr{A}( \mathscr{G}_d )[ n ] | $ that is isomorphic (as in Definition~\ref{defMAGandgraphsisomorphisms}) to $ \mathscr{G}_d $.
\end{thmafterlemmaundersection}

As an immediate corollary of Theorem~\ref{thmMAGisomorphism}, we have that the same holds for the undirected case. To achieve a simple proof of that in Corollary~\ref{corClassicMAGisomorphism}, just note that any undirected MAG (or graph) without self-loops can be equivalently represented by a directed MAG (or graph, respectively) in which, for every oriented edge (i.e., arrow), there must be an oriented edge in the exact opposite direction.
In other words, the adjacency matrix must be symmetric with respect to the main diagonal.

\begin{corollaryafterlemmaundersection}\label{corClassicMAGisomorphism}
	For every simple MAG $ \mathscr{G}_c $ (as in Definition~\ref{defSimplifiedMAG}) of order $p>0$, where all aspects are non-empty sets, there is a unique (up to a graph isomorphism) classical graph $ G_{ \mathscr{G}_c } = \left( V , E \right) $ with $ | V( G ) | =  \prod\limits_{ n = 1 }^{ p } | \mathscr{A}( \mathscr{G}_c )[ n ] | $ that is isomorphic to $ \mathscr{G}_c $.
\end{corollaryafterlemmaundersection}

From these results, we also have that the concepts of \emph{walk}, \emph{trail}, and \emph{path} become well-defined for MAGs analogously to within the context of graphs. For this purpose, see section 3.5 in \cite{Wehmuth2016b}.

\subsubsection{Algorithmic information theory}\label{subsubsectionAIT}

We now remember some important relations in algorithmic information theory \cite{Li1997,Chaitin2004,Downey2010,Calude2009,Grunwald2008} that we will apply in the proofs contained in the present article. 
More specifically, the following results can be found in \cite{Li1997,Downey2010,Chaitin2004,Calude2002}.

\begin{lemmaundersection}\label{lemmaBasicAIT}
	For every $ x , y \in \{ 0 , 1 \}^* $ and $  n \in \mathbb{N} $,
	\begin{align}
	\label{lemmaBasicAIT1} C( x ) & \leq l( x ) + \mathbf{O}(1) \\
	\label{lemmaBasicAIT10} K( x ) & \leq l( x ) + \mathbf{O}( \lg( l( x ) ) ) \\
	\label{lemmaBasicAIT2} C( y \, | x ) & \leq C( y ) + \mathbf{O}(1) \\
	\label{lemmaBasicAIT11} K( y \, | x ) & \leq K( y ) + \mathbf{O}(1) \\
	\label{lemmaBasicAIT3} C( y \, | \, x ) \leq K( y \, | \, x ) + \mathbf{O}(1) & \leq C( y \, | \, x ) + \mathbf{O}( \lg( C( y \, | \, x ) ) ) \\
	\label{lemmaBasicAIT4} C( x ) \leq C( x , y ) + \mathbf{O}(1) & \leq C( y ) + C( x \, | y ) + \mathbf{O}( \lg( C( x , y ) ) ) \\
	\label{lemmaBasicAIT5} K( x ) \leq K( x , y ) + \mathbf{O}(1) & \leq  \; K( y ) + K( x \, | y ) + \mathbf{O}(1) \\
	\label{lemmaBasicAIT6} C( x ) & \leq K( x ) + \mathbf{O}(1) \\
	\label{lemmaBasicAIT7} K( n ) & = \mathbf{O}( \lg( n ) ) \\
	\label{lemmaBasicAIT8} K( x ) & \leq C( x ) + K( C( x ) ) + \mathbf{O}(1) \\
	\label{lemmaBasicAIT9} I_A( x ; y ) & = I_A( y ; x ) \pm \mathbf{O}(1)
	\end{align}
	
%
	
\end{lemmaundersection}

Note that the inverse relation $ K( x , y ) + \mathbf{O}(1) \geq  \; K( y ) + K( x \, | y ) + \mathbf{O}(1)  $ does not hold in general in Equation~\eqref{lemmaBasicAIT5}. 
In fact, one can show that $ K( x , y ) = K( y ) + K( x \, \mid \, \left< y , K( y ) \right>  ) \pm \mathbf{O}(1) $, which is the key step to prove Equation~\eqref{lemmaBasicAIT9}.
In the present article, wherever the concept of information is mentioned, we are in fact referring to algorithmic information.

\begin{lemmaundersection}\label{lemmaComputablefunctioninK}
	Let $ \myfunc{ f_c }{ \mathbb{N} }{ \mathbb{N} }{ n }{ f_c( n ) } $ be a computable function, then
	\[
	K( f_c( n ) ) \leq K( n ) + \mathbf{O}(1) 
	\]
\end{lemmaundersection}

One of the most important results in algorithmic information theory is the investigation and proper formalization of a mathematical theory for randomness \cite{Downey2019,Calude2002}. 
For example, one of these mathematical objects satisfying this notion comes from semi-measuring the set of halting programs of a prefix-free universal programming language:

\begin{definitionafternotationundersubsection}\label{defOmeganumber}
	Let $ \Omega \in \left[ 0 , 1 \right]  \subset \mathbb{R} $ denote the \emph{halting probability} (also known as Chaitin's constant or Omega number).  The halting probability is defined by
	\[
	\Omega = \sum\limits_{ \substack{ \exists y \left( \mathbf{U}( \mathrm{p} ) = y \right) \\  \mathrm{p} \in \mathbf{L'_U} } } \quad \frac{ 1 }{ 2^{ l( \mathrm{p} ) } }
	\]
	
\end{definitionafternotationundersubsection}

This is a widely known example of infinite binary sequence, or real number, that is algorithmically random with respect to prefix algorithmic complexity:

\begin{thmafterlemmaundersection}\label{thmOmeganumber}
	Let $ n \in \mathbb{N} $.
	Then,
	\[
	K( \Omega \upharpoonright_n ) \geq n - \mathbf{O}( 1 )
	\]
	\noindent 
\end{thmafterlemmaundersection}

That is, $ \Omega $ is $ \mathbf{O}(1) $-K-random.
And, for real numbers or infinite sequences in general, one can also connect prefix algorithmic randomness with the plain algorithmic complexity of initial segments \cite{Downey2010}:

\begin{thmafterlemmaundersection}\label{thmK-randomandC-random}
	Let $ x \in \left[ 0 , 1 \right] \subset \mathbb{R} $ be a real number. Then, the following are equivalent:
	\begin{align}
	\text{ $x$ is $ \mathbf{O}(1) $-K-random } \\
	C( x \upharpoonright_n ) \geq n - K( n ) - \mathbf{O}(1) \\
	C( x \upharpoonright_n \, | \, n ) \geq n - K( n ) - \mathbf{O}(1)
	\end{align}
\end{thmafterlemmaundersection}

\subsubsection{Algorithmically random graphs}\label{subsubsectionARG}

The application of algorithmic randomness and algorithmic complexity to graph theory generated fruitful lemmas and theorems with the purpose of studying graph topological properties, such as diameter, connectivity, degree, statistics of subgraphs, unlabeled graphs counting, and automorphisms \cite{Zenil2018a,Buhrman1999} and studying algorithmic randomness of infinite graphs as model-theoretic infinite structures \cite{Khoussainov2014,Harrison-Trainor2019}. 
In this section, we restate some of these results for algorithmically random string-based representations of (finite) classical graphs in \cite{Li1997,Buhrman1999}.

\begin{lemmaundersection}\label{lemmaFractionofrandomgraphs}
	A fraction of at least $ 1 - \frac{1}{2^{ \delta(n) }} $ of all classical graphs $G$ with $ | V(G) | = n $ is $ \delta(n) $-C-random.
\end{lemmaundersection}

\begin{lemmaundersection}\label{lemmaDegreerandomgraph}
	The degree $ \mathbf{d}( v ) $ of a vertex $ v \in V(G) $ in a $ \delta(n) $-C-random classical graph $G$ with $ | V(G) | = n $ satisfies
	\[
	\left| \mathbf{d}( v ) - \left( \frac{ n - 1 }{ 2 } \right) \right| = \mathbf{O}\left( \sqrt{ n \, \left( \delta(n) + \lg(n) \right) } \right)
	\text{ .}
	\]
\end{lemmaundersection}

\begin{lemmaundersection}\label{lemmaDiameterrandomgraph}
	All $ \mathbf{o}( n ) $-C-random classical graphs $G$ with $ | V(G) | = n $ have 
	\[ \frac{n}{4} \pm \mathbf{o}(n) \]
	disjoint paths of length 2 between each pair of vertices $ u , v \in V(G) $. In particular, all $ \mathbf{o}( n ) $-C-random classical graphs $G$ with $ | V(G) | = n $ have diameter $2$.
\end{lemmaundersection}

\begin{lemmaundersection}\label{lemmaStarinrandomgraphs}
	Let $ c \in \mathbb{N} $ be a fixed constant.
	Let $ G $ be a $ ( c \, \lg(n) )$-C-random classical graph  with $ | V(G) | = n $.
	Let $ X_{ f( n ) }( v ) $ denote the set of the least $ f(n) $ neighbors of a vertex $ v \in V(G) $, where
	\[
	\myfunc{ f }{ \mathbb{N}}{ \mathbb{N} }{ n }{ f(n) }
	\text{ .}
	\]
	Then, for every vertices $ u , v \in V(G) $, either
	\begin{align*}
	\{ u , v \} & \in E(G) \\
	& \text{or} \\
	\exists i \in V(G) ( i  \in  X_{ f(n) }( v ) \land \{ u , i \} & \in E(G) \land \{ i , v \} \in E(G) ) 
	\end{align*}
	\noindent with $ f(n) \geq ( c + 3 )\lg( n )  $.
\end{lemmaundersection}

\begin{lemmaundersection}\label{lemmaRigidgraphs}
	If
	\[
	\delta( n ) = \mathbf{o}\left( n - \lg( n ) \right)
	\text{ ,}
	\]
	\noindent then all $ \delta( n ) $-C-random classical graphs are rigid.
	
\end{lemmaundersection}

\section{Recursively labeled multiaspect graphs}\label{sectionRecursivelylabeledMAGs}

In this section, we will introduce a string-based representation of a multiaspect graph (MAG). 
First, we need to generalize the concept of a labeled graph in order to grasp the set of composite vertices. 
As with labeled graphs \cite{Li1997,Buhrman1999,Zenil2018a,Zenil2014,Mowshowitz2012}, where there is an enumeration of its vertices assigning a natural number to each one of them, we want that the composite edge set $ \mathscr{E} $ continues to be uniquely represented by a binary string, except for an automorphism from composite vertices permutation or a reordering (or re-indexing) of composite edges. 
In fact, we will achieve a more general condition than a fixed lexicographical ordering of the $ | \mathbb{E}_c ( \mathscr{G}_c ) | $ edges, so that, for each multidimensional space (i.e., more formally, for each companion tuple), any MAG within this multidimensional space shares the same computable way of indexing the composite edges.
Although the recursive ordering may depend on the choice of the individual MAG with its associated companion tuple, once it is achieved, any other MAG with the same companion tuple can be univocally represented by following the same ordering,\footnote{ See Definition~\ref{BdefPlaincomplexityofedgesets}.}
but one cannot necessarily guarantee that same will hold for MAGs with distinct companion tuples.
Thus, as reinforced by the results in \cite{Abrahao2020cpublished,Abrahao2021publishednat}, Definition~\ref{defLabeledMAG} introduces MAGs that are \emph{encodable}, that is recursively labeled, in such an ``individual'' sense, which is agnostic with respect to e.g. whether one is taking two distinct recursive labelings for two distinct MAGs.

In a general sense, we say that a MAG $ \mathscr{G}_c $ from Definition~\ref{defSimplifiedMAG} is recursively labeled if and only if there is an algorithm that, given the companion tuple $ \tau( \mathscr{G}_c ) $ (see Definition~\ref{defCompaniontuple}) as input, returns a recursive bijective ordering of composite edges $ e \in \mathbb{E}_c( \mathscr{G}_c ) $. More formally:

\begin{definitionundersection}\label{defLabeledMAG}
	A MAG $ \mathscr{G}_c $ (as in Definition~\ref{defSimplifiedMAG}) is \emph{recursively labeled} given $ \tau( \mathscr{G}_c ) $  \textit{iff} there are programs $ \mathrm{p}_1 , \mathrm{p}_2  \in \{ 0 , 1 \}^* $ such that, for every $ \tau( \mathscr{G}_c ) $ with $ a_i , b_i , j \in \mathbb{N} $ and $ 1 \leq i \leq p \in \mathbb{N} $, we have that following hold at the same time:
	\begin{enumerate}[label=(\Roman{*})]
		\item\label{defLabeledMAG1} if $ \left( a_1 , \dots , a_p \right) , \left( b_1 , \dots , b_p \right) \in \mathbb{V}\left( \mathscr{G}_c \right)$, then
		\begin{align*}
		& \mathbf{U}\left( \left< \left< a_1 , \dots , a_p \right> , \left< b_1 , \dots , b_p \right> , \left< \left<  \tau( \mathscr{G}_c ) \right> , \mathrm{p}_1 \right>  \right> \right)  = { \left( j \right) }_2
		\end{align*}
		
		\item\label{defLabeledMAG1.2} if $  \left( a_1 , \dots , a_p \right)  $ or $ \left( b_1 , \dots , b_p \right) $ does not belong to $ \mathbb{V}\left( \mathscr{G}_c \right) $, then
		\begin{align*}
		& \mathbf{U}\left( \left< \left< a_1 , \dots , a_p \right> , \left< b_1 , \dots , b_p \right> , \left< \left<  \tau( \mathscr{G}_c ) \right> , \mathrm{p}_1 \right>  \right> \right)  = 0
		\end{align*}
		
		\item\label{defLabeledMAG2} if \[ 1 \leq j \leq \left| \mathbb{E}_c( \mathscr{G}_c )   \right| = \frac{ { \left| \mathbb{V}( \mathscr{G}_c ) \right| }^2 - { \left| \mathbb{V}( \mathscr{G}_c ) \right| } }{ 2 }  \text{ ,}\] then
		\begin{align*}
		& \mathbf{U}\left( \left<  j  , \left< \left< \tau( \mathscr{G}_c ) \right> , \mathrm{p}_2 \right>  \right>  \right) = \left< \left< a_1 , \dots , a_p  \right> , \left<  b_1 , \dots , b_p \right>  \right> = { \left( e_j \right) }_2
		\end{align*}
		
		\item\label{defLabeledMAG2.2} if \[ 1 \leq j \leq \left| \mathbb{E}_c( \mathscr{G}_c )   \right| = \frac{ { \left| \mathbb{V}( \mathscr{G}_c ) \right| }^2 - { \left| \mathbb{V}( \mathscr{G}_c ) \right| } }{ 2 }  \] does not hold, then
		\begin{align*}
		& \mathbf{U}\left( \left<  j  , \left< \left< \tau( \mathscr{G}_c ) \right> , \mathrm{p}_2 \right>  \right>  \right) = \left< \left< a_1 , \dots , a_p  \right> , \left<  b_1 , \dots , b_p \right>  \right> = \left< 0 \right>
		\end{align*}
	\end{enumerate}
\end{definitionundersection}

Besides simple MAGs, note that this Definition~\ref{defLabeledMAG} can be easily extended to arbitrary MAGs as in Definition~\ref{defMAG}.

We can show that Definition~\ref{defLabeledMAG} is always satisfiable by MAGs that have every element of its aspects labeled as a natural number: 

\begin{lemmaundersection}\label{lemmaLabeledMAG}
	Any arbitrary simple MAG $ \mathscr{G}_c $  with $ \mathscr{A}(\mathscr{G}_c )[ i ] = \{ 1, \dots , \left| \mathscr{A}(\mathscr{G}_c )[ i ] \right| \} \subset \mathbb{N} $, where $ \left| \mathscr{A}( {\mathscr{G}_c} )[ i ] \right| \in \mathbb{N} $ and $ 1 \leq i \leq p = \left|\mathscr{A}( {\mathscr{G}_c} )  \right| \in \mathbb{N} $, is recursively labeled given $ \tau( \mathscr{G}_c ) $ (i.e., it satisfies Definition~\ref{defLabeledMAG}).
	
	\begin{proof}
		Since $ \left< \cdot , \cdot \right> $ represents a recursive bijective pairing function, the companion tuple 
		\[  \left< \tau( \mathscr{G}_c ) \right> = \left< | \mathscr{A}( \mathscr{G}_c )[1] |, \dots , | \mathscr{A}( \mathscr{G}_c )[p] | \right>  \] 
		univocally determines the value of $ p $ and the maximum value for each aspect.\footnote{ See also \cite{Wehmuth2017} for more properties of the companion tuple regarding generalized graph representations. } 
		Hence, given any  $  \left< \tau( \mathscr{G}_c ) \right> $, one can always define a recursive lexicographical ordering $ <_{ \mathbb{V} } $ of the set $ \{ \left< x_1 , \dots , x_p  \right> \, \mid \, ( x_1 , \dots , x_p )  \in \mathbb{V}( \mathscr{G}_c ) \} $ by: starting at $ \left< 1 , \dots , 1 \right> $ and; from a recursive iteration of this procedure from the right character to the left character, ordering all possible arrangements of the rightmost characters while one maintains the leftmost characters fixed, while respecting the limitations $ | \mathscr{A}( \mathscr{G}_c )[i]  | $ with $ 1 \leq i \leq p \in \mathbb{N} $ in each $ \mathscr{A}(\mathscr{G}_c )[ i ] $. 
		That is, from choosing an arbitrary well-known lexicographical ordering of ordered pairs, one can iterate this for a lexicographical ordering of $n$-tuples by $ \left( x_1 , \left( x_2 , x_3 \right) \right) = \left( x_1 ,  x_2 , x_3 \right)  $, $ \left( x_1 , \left(  x_2 , \left( x_3 , x_4 \right) \right) \right) = \left( x_1 ,  x_2 , x_3 , x_4 \right)  $, and so on.\footnote{ See also Definition~\ref{notationPairing}. }---alternatively, one may contruct the order relation $ <_{ \mathbb{V} } $ from functions $ D\left( \mathbf{ u } , \tau \right) $ and $ N\left( d , i , \tau \right) $ defined in \cite[Section 3.2: Ordering of Composite Vertices and Aspects, p. 9]{Wehmuth2017}.
		Therefore, from this recursive bijective ordering of composite vertices (given any  $  \left< \tau( \mathscr{G}_c ) \right> $), we will now construct a sequence defined by a recursive bijective ordering $ <_{ \mathbb{E}_c } $ of the composite edges of a MAG $  \mathscr{G}_c  $. To this end, from the order relation $ <_{ \mathbb{V} } $, one first build a sequence by applying a classical lexicographical ordering $ <_{ \mathbb{E} } $ to the set of pairs
		\[
		\big\{ \left< \left< x_1 , \dots , x_p \right> ,   \left< y_1 , \dots , y_p \right>  \right>   \, \big\mid \, ( x_1 , \dots , x_p ) ,  ( y_1 , \dots , y_p ) \in \mathbb{V}( \mathscr{G}_c ) \big\} 
		\]
		\noindent Then, one excludes the occurrence of self-loops and the second occurrence of symmetric pairs 
		
		\noindent $ \left( \left<  y_1 , \dots , y_p \right>  , \left< x_1 , \dots , x_p  \right> \right) $, generating a subsequence of the previous sequence. Note that the procedure for determining whether the two composites vertices in an composite edge are equal or not is always decidable, so that self-loops on composite vertices will not return index values under order relation $ <_{ \mathbb{E}_c } $. Additionally, note that, since the sequence of composites edges was formerly arranged in lexicographical order relation $ <_{ \mathbb{E} } $, then, for every $ a , b \in \mathbb{N} $ under order relation $ <_{ \mathbb{V} } $,
		\[
		a <_{ \mathbb{V} } b \implies ( a , b )  <_{ \mathbb{E} } ( b , a )
		\]  
		\noindent This way, since subsequences preserve order, if $ i_{ ( a , b )_{ <_{ \mathbb{E} } } } $ is the index value of the pair $ ( a , b ) $ in the sequence built under order relation $ <_{ \mathbb{E} } $ and $ a <_{ \mathbb{V} } b $, then 
		\[
		i_{ ( b , a )_{ <_{ \mathbb{E}_c } } } \coloneq i_{ ( a , b )_{ <_{ \mathbb{E} } } }  
		\text{ and } 
		i_{ ( a , b )_{ <_{ \mathbb{E}_c } } } \coloneq i_{ ( a , b )_{ <_{ \mathbb{E} } } }
		\]
		
		\noindent Thus, let $  \mathrm{p}_1 $ be a fixed string that represents on a universal Turing machine the algorithm  that, given $ \tau( \mathscr{G}_c ) $, $ \left< a_1 , \dots , a_p \right> $, and $ \left< b_1 , \dots , b_p \right> $ as inputs, 
		\begin{enumerate}[label=(\roman*)]
			\item builds the sequence of composite edges by the order relation $ <_{ \mathbb{E}_c } $ described before such that, for each step of this construction, 
			\begin{enumerate}
				\item search for $ \left( \left( a_1 , \dots , a_p \right) , \left( b_1 , \dots , b_p \right) \right) $ or $ \left( \left( b_1 , \dots , b_p \right) , \left( a_1 , \dots , a_p \right) \right) $ in this sequence;
				
				\item if one of these pairs is found, it returns the index value of the first one of these pairs found in this sequence;
				
				\item else, it continues building the sequence;
				
			\end{enumerate}
			\item if the sequence is completed\footnote{ Note that $ \mathbb{V}\left( \mathscr{G}_c \right) $ is always finite.} and neither 
			\[ \left( \left( a_1 , \dots , a_p \right) , \left( b_1 , \dots , b_p \right) \right) \] nor \[ \left( \left( b_1 , \dots , b_p \right) , \left( a_1 , \dots , a_p \right) \right) \] were found, then returns $ 0  $.
		\end{enumerate} 
		Note that, if $ \left( a_1 , \dots , a_p \right) , \left( b_1 , \dots , b_p \right) \in \mathbb{V}\left( \mathscr{G}_c \right)$, then one of these pairs must be always eventually found, since $ \left| \mathscr{A}( {\mathscr{G}_c} )[ i ] \right| \in \mathbb{N} $ with $ 1 \leq i \leq p = \left|\mathscr{A}( {\mathscr{G}_c} )  \right| \in \mathbb{N} $.
		An analogous algorithm defines $  \mathrm{p}_2  $, but by searching for the $j$-th element in the sequence generated by the order relation $ <_{ \mathbb{E}_c } $ and returning the respective pair of tuples instead (or $ \left< 0 \right> $, if $1 \leq j \leq \left| \mathbb{E}_c( \mathscr{G}_c )   \right| = \frac{ { \left| \mathbb{V}( \mathscr{G}_c ) \right| }^2 - { \left| \mathbb{V}( \mathscr{G}_c ) \right| } }{ 2 }  $ does not hold).
	\end{proof}
\end{lemmaundersection}


The reader may note that, unlike the recursive labeling of Definition~\ref{defLabeledMAG},  we will later on demonstrate in Lemma~\ref{lemmaLabeledfamilyofMAG} with Definition~\ref{defLabeledfamilyofMAG}, that there is a unified way to index (or order) composite edges. 
And this ``collective'' or ``global'' recursive labeling will hold for every finite MAG that have the same number of equal-size aspects. 

In any event, with this pair of programs $ \mathrm{p}_1 , \mathrm{p}_2 $, and with $ \left<  \tau( \mathscr{G}_c ) \right> $, one can always build an algorithm that, given a bit string $ x \in \{ 0 , 1 \}^* $ of length $ \left| \mathbb{E}_c( \mathscr{G}_c )   \right|  $ as input, computes a composite edge set $ \mathscr{E}( \mathscr{G}_c ) $ and build another algorithm that, given the composite edge set $ \mathscr{E}( \mathscr{G}_c ) $ as input, returns a string $x$.
Therefore, there are strings $x$ that determine (up to an automorphism from composite vertices permutation or up to a reordering of composite edges) the presence or absence of composite edges for the recursively labeled (finite) MAG $ \mathscr{G}_c  $. 
This is a directly analogous phenomenon previously studied in \cite{Li1997,Buhrman1999} for algorithmically random finite classical graphs and in \cite{Khoussainov2014} for infinite graphs.
We call such a string a \emph{characteristic string} of the MAG.

In fact, one can define characteristic strings in an agnostic manner regarding any encoding or recursive indexing of (composite) edges:
\begin{definitionundersection}\label{BdefCharacteristicstringofasimpleMAG}
	Let $ \left( e_1 , \dots , e_{ \left| \mathbb{E}_c( \mathscr{G}_c )   \right| } \right) $ be any arbitrarily fixed ordering of all possible composite edges of a  simple MAG $ \mathscr{G}_c $ as in Notations~\ref{BdefPlaincomplexityofedgesets} and ~\ref{BdefPrefixcomplexityofedgesets}.
	We say that a string $ x \in \{ 0 , 1 \}^* $ with $ l( x ) = \left| \mathbb{E}_c( \mathscr{G}_c )   \right| $ is a \emph{characteristic string} of a simple MAG $ \mathscr{G}_c $
	\textit{iff}, for every $ e_j \in \mathbb{E}_c( \mathscr{G}_c ) $,
	\[
	e_j  \in \mathscr{E}( \mathscr{G}_c  ) \iff \text{ the $j$-th digit in $x$ is $1$}
	\text{ ,}
	\]
	\noindent  where $ 1 \leq j \leq l( x ) $. 
	
\end{definitionundersection}

Thus, if the ordering assumed in Definition~\ref{BdefCharacteristicstringofasimpleMAG} matches the same ordering embedded in the ordering given by programs $ \mathrm{p}_1 , \mathrm{p}_2 $, we will show now in Lemma~\ref{lemmaBasicMAGandstrings} that both the MAG and its respective characteristic string are equivalent in terms of algorithmic information, except for the minimum information necessary to encode the multidimensional space.
However, as we will see later on, this condition becomes unnecessary for recursively labeled families, since this ordering is already embedded in Definition~\ref{defLabeledfamilyofMAG}, holding for the entire family of every possible MAG with the same order and equal-size aspects.
See also Lemma~\ref{lemmaLabeledfamilyofMAG}.
For the present purposes, Definitions~\ref{defLabeledMAG} and \ref{BdefCharacteristicstringofasimpleMAG} give rise to the following Lemma:

\begin{lemmaundersection}\label{lemmaBasicMAGandstrings}
	Let $ x \in \{ 0 , 1 \}^* $.
	Let $ \mathscr{G}_c  $ be a recursively labeled MAG given $ \tau( \mathscr{G}_c ) $ (as in Definition~\ref{defLabeledMAG}) 
	such that $ x $ is the respective characteristic string of the recursive labeling of $ \mathscr{G}_c  $ (as in Definition~\ref{BdefCharacteristicstringofasimpleMAG}).
	\noindent  where $ 1 \leq j \leq l( x ) $.
	Then,
	\begin{align}
	\label{lemmaBasicMAGandstrings1} C( \mathscr{E}( \mathscr{G}_c  ) \, | \, x  ) \leq K( \mathscr{E}( \mathscr{G}_c  ) \, | \, x  ) + \mathbf{O}(1)  & =  K( \left< \tau( \mathscr{G}_c ) \right> ) + \mathbf{O}(1)  \\
	\label{lemmaBasicMAGandstrings2} C( x \, | \, \mathscr{E}( \mathscr{G}_c  )  )  \leq K( x \, | \, \mathscr{E}( \mathscr{G}_c  )  ) + \mathbf{O}(1)  & =  K( \left< \tau( \mathscr{G}_c ) \right> ) + \mathbf{O}(1) \\
	\label{lemmaBasicMAGandstrings3} K( x ) & = K( \mathscr{E}( \mathscr{G}_c  ) )  \pm \mathbf{O}\left( K( \left< \tau( \mathscr{G}_c ) \right> ) \right) \\
	\label{lemmaBasicMAGandstrings4} I_A( x ; \mathscr{E}( \mathscr{G}_c  ) ) = I_A( \mathscr{E}( \mathscr{G}_c  ) ; x ) \pm \mathbf{O}(1) & = K( x ) - \mathbf{O}\left( K( \left< \tau( \mathscr{G}_c ) \right> )  \right) 
	\end{align}
	\noindent \\
	
	\begin{proof}[Proofs]
		\noindent \\
		\begin{enumerate}
			\item[(proof of \ref{lemmaBasicMAGandstrings1})]\label{prooflemmaBasicMAGandstrings1} First, remember notation of $ \mathscr{E}( \mathscr{G}  ) $ in Definitions~\ref{BdefK} and~\ref{BdefC} from which we have that
			\[
			K( \left< \mathscr{E}( \mathscr{G}  ) \right> ) = K( \left<  \left< e_1, z_1 \right>, \dots , \left< e_n , z_n \right>  \right> )
			\]
			\noindent such that 
			\[
			z_i = 1  \iff e_i \in \mathscr{E}( \mathscr{G} )
			\text{ ,}
			\]
			\noindent where  $ z_i \in \{ 0 , 1 \} $ with $ 1 \leq i \leq n= | \mathbb{E}( \mathscr{G} ) | $. Thus, for MAGs $ \mathscr{G}_c $ defined in Definition~\ref{defSimplifiedMAG}, we will have that
			\[
			K( \left< \mathscr{E}( \mathscr{G}_c  ) \right> ) = K( \left<  \left< e_1, z_1 \right>, \dots , \left< e_n , z_n \right>  \right> )
			\]
			\noindent such that 
			\[
			z_i = 1  \iff e_i \in \mathscr{E}( \mathscr{G} )
			\text{ ,}
			\]
			\noindent where  $ z_i \in \{ 0 , 1 \} $ with 
			\[
			1 \leq i \leq n= | \mathbb{E}_c( \mathscr{G}_c ) | = \frac{ { \left| \mathbb{V}( \mathscr{G}_c ) \right| }^2 - { \left| \mathbb{V}( \mathscr{G}_c ) \right| } }{ 2 }
			\]
			\noindent \\
			We also have that, since $ \mathscr{G}_c  $ is a recursively labeled MAG, there is $ \mathrm{p}_2 $ such that Definition~\ref{defLabeledMAG} holds independently of the chosen companion tuple $ \tau( \mathscr{G}_c ) $. Let $ \left< \tau( \mathscr{G}_c ) \right> $ be a self-delimiting string that encodes the companion tuple $ \tau( \mathscr{G}_c ) $. Let $ p $ be a binary string that represents on a universal Turing machine the algorithm that reads the companion tuple $  \left< \tau( \mathscr{G}_c ) \right> $ as his first input and reads the string $x$ as its second input.
			Then, it reads each $j$-th bit of $x$, runs $   \left<  j  , \left< \left< \tau( \mathscr{G}_c ) \right> , \mathrm{p}_2 \right>  \right>  $ and, from the outputs $ e_j $ of these programs $   \left<  j  , \left< \left< \tau( \mathscr{G}_c ) \right> , \mathrm{p}_2 \right>  \right>  $, returns the string $ \left<  \left< e_1, z_1 \right>, \dots , \left< e_n , z_n \right>  \right> $ where $ z_j = 1 $, if the $j$-th bit of $x$ is $1$, and $ z_j = 0 $, if the $j$-th bit of $x$ is $0$. Thus, we will have that there is a binary string $ \left< \left< \tau( \mathscr{G}_c ) \right> ,  p \right> \in \mathbf{L'_U}$ that represents an algorithm running on a prefix (or self-delimiting) universal Turing machine $ \mathbf{U} $ that, given $x$ as input, runs $p$ taking also $ \left< \tau( \mathscr{G}_c ) \right> $ into account. Since $ \mathrm{p}_2 $ is fixed, we will have that there is a self-delimiting binary encoding of $ \left( \left< \tau( \mathscr{G}_c ) \right> ,  p \right) $ with
			\[
			l\left( \left< \left< \tau( \mathscr{G}_c ) \right> ,  p \right> \right) \leq K( \left< \tau( \mathscr{G}_c ) \right> ) + \mathbf{O}(1)
			\] 
			\noindent Then, by the minimality of $ K( \cdot ) $, we will have that
			\[
			K( \mathscr{E}( \mathscr{G}_c  ) \, | \, x  ) \leq  	l\left( \left< \left< \tau( \mathscr{G}_c ) \right> ,  p \right> \right) \leq K( \left< \tau( \mathscr{G}_c ) \right> ) + \mathbf{O}(1)
			\]
			\noindent The inequality $ C( \mathscr{E}( \mathscr{G}_c  ) \, | \, x  ) \leq K( \mathscr{E}( \mathscr{G}_c  ) \, | \, x  ) + \mathbf{O}(1)  $ follows directly from Lemma~\ref{lemmaBasicAIT}. \\
			
			\item[(proof of \ref{lemmaBasicMAGandstrings2})]\label{prooflemmaBasicMAGandstrings2}
			This proof follows analogously to the proof of Equation~\eqref{lemmaBasicMAGandstrings1}, but using program $ \mathrm{p}_1 $ instead of $ \mathrm{p}_2 $ in order to build the string $x$ from $ \left< \mathscr{E}( \mathscr{G}_c  ) \right>  $. \\
			
			\item[(proof of \ref{lemmaBasicMAGandstrings3})]\label{prooflemmaBasicMAGandstrings3} This proof follows analogously to the proof of Equation \ref{lemmaBasicAIT5} in Lemma \ref{lemmaBasicAIT}. Let $p$ be a shortest self-delimiting description of $ \left< \mathscr{E}( \mathscr{G}_c  ) \right> $. From Equation \ref{lemmaBasicMAGandstrings2}, we know there is $ q $, independent of the choice of $p$, such that it is a shortest self-delimiting description of $x$ given $ \left< \mathscr{E}( \mathscr{G}_c  ) \right>  $, where 
			\[
			K( x \, | \, \mathscr{E}( \mathscr{G}_c  )  )  =  K( \left< \tau( \mathscr{G}_c ) \right> ) + \mathbf{O}(1)
			\]
			\noindent Thus, there is string $ s $, independent of the choice of $p$ and $q$, that represents the algorithm running on a universal Turing machine that, given $p$ and $q$ as its inputs, calculates the output of $p$, and returns the output of running $q$ with the output of $p$ as its input. We will have that there is a prefix universal machine $ \mathbf{U} $ in which $ \left< p , q , s \right> \in \mathbf{L'_U} $ and, from Equation \ref{lemmaBasicMAGandstrings2},
			\begin{align*}
			| \left< p , q , s \right> | & \leq K( \mathscr{E}( \mathscr{G}_c  ) ) + K( x \, | \, \mathscr{E}( \mathscr{G}_c  )  ) +  \mathbf{O}(1) \leq \\
			& \leq K( \mathscr{E}( \mathscr{G}_c  ) )  + K( \left< \tau( \mathscr{G}_c ) \right> ) + \mathbf{O}(1)
			\end{align*}
			\noindent Then, by the minimality of $ K( \cdot ) $, we will have that
			\[
			K( x ) \leq | \left< p , q , s \right> | \leq K( \mathscr{E}( \mathscr{G}_c  ) )  + K( \left< \tau( \mathscr{G}_c ) \right> ) + \mathbf{O}(1)
			\]
			\noindent Therefore, 
			\[
			K( x ) \leq K( \mathscr{E}( \mathscr{G}_c  ) )  + \mathbf{O}\left( K( \left< \tau( \mathscr{G}_c ) \right> ) \right)
			\]
			\noindent The proof of $ K( \mathscr{E}( \mathscr{G}_c  ) ) \leq K( x )  + K( \left< \tau( \mathscr{G}_c ) \right> ) + \mathbf{O}(1) $ follows in the same manner, but using Equation~\eqref{lemmaBasicMAGandstrings1} instead of \ref{lemmaBasicMAGandstrings2}. \\
			
			\item[(proof of \ref{lemmaBasicMAGandstrings4})]\label{prooflemmaBasicMAGandstrings4}
			We have from Definition~\ref{BdefK} that 
			\begin{equation}\label{stepI_AforMAGSandx}
			I_A( x \, ; \mathscr{E}( \mathscr{G}_c  ) ) = K( \mathscr{E}( \mathscr{G}_c  ) ) - K( \mathscr{E}( \mathscr{G}_c  ) \, | x^* ) 
			\end{equation}
			\noindent Now, we build a program for $ \mathscr{E}( \mathscr{G}_c  )  $ given $  x^* $ almost identical to the one in the proof of Equation~\eqref{lemmaBasicMAGandstrings1}. First, remember that, since $ \mathscr{G}_c  $ is a recursively labeled MAG, there is $ \mathrm{p}_2 $ such that Definition~\ref{defLabeledMAG} holds independently of the chosen companion tuple $ \tau( \mathscr{G}_c ) $. Let $ \left< \tau( \mathscr{G}_c ) \right> $ be a self-delimiting string that encodes the companion tuple $ \tau( \mathscr{G}_c ) $. Let $ p $ be a binary string that represents the algorithm running on a universal Turing machine that reads the companion tuple $  \left< \tau( \mathscr{G}_c ) \right> $ as his first input and reads the output of $ x^* $ as its second input. Then, it reads each $j$-th bit of\footnote{ Note that $x$ is the output of $ x^* $ on the chosen universal Turing machine. } $x$, runs $   \left<  j  , \left< \left< \tau( \mathscr{G}_c ) \right> , \mathrm{p}_2 \right>  \right>  $ and, from the outputs of $   \left<  j  , \left< \left< \tau( \mathscr{G}_c ) \right> , \mathrm{p}_2 \right>  \right>  $, returns the string $ \left<  \left< e_1, z_1 \right>, \dots , \left< e_n , z_n \right>  \right> $ where $ z_j = 1 $, if the $j$-th bit of $x$ is $1$, and $ z_j = 0 $, if the $j$-th bit of $x$ is $0$. Therefore, we will have that there is a binary string $ \left< \left< \tau( \mathscr{G}_c ) \right> ,  p \right> \in \mathbf{L'_U}$ that represents an algorithm running on a prefix (or self-delimiting) universal Turing machine $ \mathbf{U} $ that, given $x^*$ as input, runs $p$ taking also $ \left< \tau( \mathscr{G}_c ) \right> $ into account. Since $ \mathrm{p}_2 $ is fixed, we have that there is a self-delimiting binary encoding of $ \left( \left< \tau( \mathscr{G}_c ) \right> ,  p \right) $ with
			\[
			l\left( \left< \left< \tau( \mathscr{G}_c ) \right> ,  p \right> \right) \leq K( \left< \tau( \mathscr{G}_c ) \right> ) + \mathbf{O}(1)
			\] 
			\noindent Then, by the minimality of $ K( \cdot ) $, we will have that
			\[
			K( \mathscr{E}( \mathscr{G}_c  ) \, | \, x^*  ) \leq  l\left( \left< \left< \tau( \mathscr{G}_c ) \right> ,  p \right> \right) \leq K( \left< \tau( \mathscr{G}_c ) \right> ) + \mathbf{O}(1)
			\]
			Thus, from Equation~\eqref{stepI_AforMAGSandx}, we will have that 
			\begin{align*}
			\mathbf{O}(1) + K( \mathscr{E}( \mathscr{G}_c  ) ) &\geq I_A( x \, ; \mathscr{E}( \mathscr{G}_c  ) ) \geq \\
			&\geq K( \mathscr{E}( \mathscr{G}_c  ) ) - \left( K( \left< \tau( \mathscr{G}_c ) \right> ) + \mathbf{O}(1) \right)
			\end{align*}
			Therefore,
			\[
			I_A( x \, ; \mathscr{E}( \mathscr{G}_c  ) ) = K( \mathscr{E}( \mathscr{G}_c  ) ) - \mathbf{O}\left( K( \left< \tau( \mathscr{G}_c ) \right> )  \right)
			\]
			For the proof of $ I_A( \mathscr{E}( \mathscr{G}_c  ) \, ; x ) = K( x ) - \mathbf{O}\left( K( \left< \tau( \mathscr{G}_c ) \right> )  \right) $, the same follows analogously to the previous proof, but using an almost identical recursive procedure to the one for Equation~\eqref{lemmaBasicMAGandstrings2} instead.
			Finally, the proof of $ I_A( x ; \mathscr{E}( \mathscr{G}_c  ) ) = I_A( \mathscr{E}( \mathscr{G}_c  ) ; x ) \pm \mathbf{O}(1) $ follows directly from Lemma~\ref{lemmaBasicAIT}.
		\end{enumerate}
	\end{proof}
\end{lemmaundersection}

The reader may notice that the results of Lemma~\ref{lemmaBasicMAGandstrings} can indeed be a corollary of the fact that the composite edge string $ \left< \mathscr{E}( \mathscr{G}_c  ) \right> $ and the correspondent characteristic string $x$ are Turing equivalent by the assumption that the MAG is already recursively labeled.
In order to facilitate future applications (and time complexity analyses) of the results in this article to algorithmic-information-based algorithms for measuring multidimensional network complexity, we choose to more explicitly address each procedure in the proof of Lemma~\ref{lemmaBasicMAGandstrings}.

Thus, for the present purposes, we have that the notion of \emph{network topological (algorithmic) information} of data representations of a finite simple MAG $ \mathscr{G}_c $, i.e., the number of bits of computably irreducible information necessary to determine/compute $ \left< \mathscr{E}( \mathscr{G}_c ) \right> $ (or to compute the set $ \mathscr{E}( \mathscr{G}_c ) $ encoded in any other data structure that is Turing equivalent to strings $ \left< \mathscr{E}( \mathscr{G}_c ) \right> $'s as e.g. in \cite{Zenil2017a,Zenil2014,Zenil2019c} for classical graphs and \cite{Santoro2019,Abrahao2019bnat} for multiplex networks), is indeed formally captured by 
\[ I_A( \left< \mathscr{E}( \mathscr{G}_c ) \right> \, ; \left< \mathscr{E}( \mathscr{G}_c ) \right> ) = K( \left< \mathscr{E}( \mathscr{G}_c ) \right> ) \pm \mathbf{O}(1) \text{ .} \]
Note that, in the present article, wherever the concept of information is mentioned, we are in fact referring to algorithmic information.

\subsection{Recursively labeled family of multiaspect graphs}\label{subsectionRecursivelylabeledfamilyofMAGs}

Another family of MAGs from Definition~\ref{defLabeledMAG} that may be of interest is the one in which the ordering of edges does not depend on $ \tau( \mathscr{G}_c ) $, i.e., on the class of aspects $ \mathscr{A}( \mathscr{G}_c ) $ and the ordering it is being encoded. 
The main idea underlying the definition of such family is that the ordering of the composite edges does not change as the companion tuple $ \tau( \mathscr{G}_c ) $ changes. 
For this purpose, we need to gather MAGs in families in which the recursively labeling does not depend on the companion tuple information.
In fact, the underlying idea is that the recursively labeling will be one of the distinctive features or characteristics, shared by each member of a family, that define such a collection of MAGs as a family.
However, it is important to remember that, in the general case, the companion tuple may be highly informative in fully characterizing the respective MAG, as we showed in \cite{Abrahao2020cpublished,Abrahao2021publishednat}.

Note that, since $ \left< \tau( \mathscr{G}_c ) \right> $ is being given as an input, it needs to be self-delimited. 
In addition, the recursive bijective pairing $ \left< \cdot  ,  \cdot \right> $ allows one to univocally retrieves the tuple $ \left( | \mathscr{A}( \mathscr{G} )[1] |, \dots , | \mathscr{A}( \mathscr{G} )[p] | \right) $. Thus, the companion tuple also informs the order of the MAG. 
Secondly, note that, for the same value of $ p = \left|  \mathscr{A}( \mathscr{G} )  \right| $ and the same value of $  | \mathbb{V}( \mathscr{G}_c )  |  $, one may have different companion tuples. Therefore, these give rise to the need of grasping the strong notion of \emph{recursive labeling} into distinct families of MAGs as follows:

\begin{definitionafternotationundersubsection}\label{defLabeledfamilyofMAG}
	A family $ F_{ \mathscr{G}_c } $ of simple MAGs $ \mathscr{G}_c $ (as in Definition~\ref{defSimplifiedMAG}) is \emph{recursively labeled} \textit{iff} there are programs $ \mathrm{p'}_1 , \mathrm{p'}_2  \in \{ 0 , 1 \}^* $ such that, for every $ \mathscr{G}_c \in F_{ \mathscr{G}_c } $ and for every $ a_i , b_i , j \in \mathbb{N} $ with $ 1 \leq i \leq p \in \mathbb{N}$, the following hold at the same time:
	
	\begin{enumerate}[label=(\Roman{*})]
		\item\label{defLabeledfamilyofMAG1} if $ \left( a_1 , \dots , a_p \right) , \left( b_1 , \dots , b_p \right) \in \mathbb{V}\left( \mathscr{G}_c \right)$, then
		\begin{align*}
		& \mathbf{U}\left( \left< \left< a_1 , \dots , a_p \right> , \left< b_1 , \dots , b_p \right> ,  \mathrm{p'}_1   \right> \right)  = { \left( j \right) }_2
		\end{align*}
		
		\item\label{defLabeledfamilyofMAG1.2} if $  \left( a_1 , \dots , a_p \right)  $ or $ \left( b_1 , \dots , b_p \right) $ does not belong to any $ \mathbb{V}\left( \mathscr{G}_c \right) $ with $ \mathscr{G}_c \in F_{ \mathscr{G}_c } $, then
		\begin{align*}
		& \mathbf{U}\left( \left< \left< a_1 , \dots , a_p \right> , \left< b_1 , \dots , b_p \right> ,  \mathrm{p'}_1   \right> \right)  = 0
		\end{align*}
		
		\item\label{defLabeledfamilyofMAG2} if \[ 1 \leq j \leq \left| \mathbb{E}_c( \mathscr{G}_c )   \right| = \frac{ { \left| \mathbb{V}( \mathscr{G}_c ) \right| }^2 - { \left| \mathbb{V}( \mathscr{G}_c ) \right| } }{ 2 }  \text{ ,}\] then
		\begin{align*}
		&\mathbf{U}\left( \left<  j  , \mathrm{p'}_2  \right>  \right) = \left< \left< a_1 , \dots , a_p  \right> , \left<  b_1 , \dots , b_p \right>  \right> = { \left( e_j \right) }_2
		\end{align*}
		
		\item\label{defLabeledfamilyofMAG.2} if \[ 1 \leq j \leq \left| \mathbb{E}_c( \mathscr{G}_c )   \right| = \frac{ { \left| \mathbb{V}( \mathscr{G}_c ) \right| }^2 - { \left| \mathbb{V}( \mathscr{G}_c ) \right| } }{ 2 }  \] does not hold for any $ \mathbb{V}\left( \mathscr{G}_c \right) $ with $ \mathscr{G}_c \in F_{ \mathscr{G}_c } $, then
		\begin{align*}
		& \mathbf{U}\left( \left<  j  , \mathrm{p'}_2  \right>  \right) = \left< \left< a_1 , \dots , a_p  \right> , \left<  b_1 , \dots , b_p \right>  \right> = \left< 0 \right>
		\end{align*}
	\end{enumerate}
	%
	
\end{definitionafternotationundersubsection}

As expected for an abstract data structure, note that $ \mathrm{p'}_1 $ and $ \mathrm{p'}_2 $ in Definition~\ref{defLabeledfamilyofMAG} define a family of finite MAGs that has an embedded unique way of indexing the composite edges for every member of the family. 
That is, there is a unique ordering for all possible composite edges in the sets $ \mathbb{E}_c( \mathscr{G}_c ) $'s such that this ordering does not depend on the choice of the MAG $ \mathscr{G}_c $ in the family $ F_{ \mathscr{G}_c } $.
Indeed, we will see in Lemma~\ref{lemmaLabeledfamilyofMAG} that, given an arbitrarily fixed order $p$, there is a recursively labeled infinite family that contains every possible MAG of order $p$ with equal-size aspects, so that such a unified ordering of composite edges is shared by each one the member of the family.
This way, those matching conditions on composite edge orderings that underlie Lemma~\ref{lemmaBasicMAGandstrings} and the results in \cite{Abrahao2020cpublished,Abrahao2021publishednat} will become redundant in Corollaries~\ref{corFamilyoflabeledMAGandstrings} and \ref{corFamilygraphsandstrings}.

The reader is also invited to note that this Definition~\ref{defLabeledfamilyofMAG} can be easily extended to arbitrary MAGs $ \mathscr{G} $, as in Definition~\ref{defMAG}. In this case, we will have $ \left| \mathbb{E}( \mathscr{G} )   \right| = { \left| \mathbb{V}( \mathscr{G} ) \right| }^2 $ instead of 
\[
\left| \mathbb{E}_c( \mathscr{G}_c )   \right| = \frac{ { \left| \mathbb{V}( \mathscr{G}_c ) \right| }^2 - { \left| \mathbb{V}( \mathscr{G}_c ) \right| } }{ 2 } 
\]

In order to show in Lemma~\ref{lemmaLabeledfamilyofMAG} that Definition~\ref{defLabeledfamilyofMAG} is satisfiable by an \emph{infinite} (recursively enumerable) family of simple MAGs, we will define an infinite family of MAGs $ \mathscr{G}_c $ such that every one of them has the same order and no condition of the presence or absence of a composite edge was taken into account.
Lemma~\ref{lemmaLabeledfamilyofMAG} ensures not only that one can follow the same ordering (or indexing) for the set $ \mathbb{E}_c( \mathscr{G}_c ) $, where this ordering does not depend on the choice of $ \mathscr{G}_c $, but also that every possible MAG of order $p$ with equal-size aspects belongs to such an infinite family satisfying Lemma~\ref{lemmaLabeledfamilyofMAG}.
That is, given an arbitrarily fixed order $p$, there is a recursively labeled (as in Definition~\ref{defLabeledfamilyofMAG}) infinite family that contains every possible MAG of order $p$ whose aspects at each iterative step have the same size. 
The key idea of the following proof is to start with an arbitrarily chosen MAG and construct an infinite family from an iteration in which the number of elements in the aspects increases uniformly. 
This way, any addition of information that the companion tuple may give, can be neutralized.
Further on, we will see in Section~\ref{subsectionNestedMAGs} that the recursively labeling developed in the proof of Lemma~\ref{lemmaLabeledfamilyofMAG} was already done in such a way that there can be a chain of nested MAGs (a fact that will be further explored in Section~\ref{subsectionNestedMAGs}).

\begin{lemmaundersection}\label{lemmaLabeledfamilyofMAG}
	There is a recursively labeled infinite family $ F_{ \mathscr{G}_c } $ of simple MAGs $ \mathscr{G}_c $ with arbitrary symmetric adjacency matrix (i.e., with arbitrary composite edge set in $ \mathbb{E}_c $). In particular, there is a recursively labeled infinite family $ F_{ \mathscr{G}_c } $ of simple MAGs $ \mathscr{G}_c $ with arbitrary symmetric adjacency matrix such that every one of them has the same arbitrary order $p$ with equal-size aspects. 
	
	\begin{proof}
		Let $ p, n_0 \in \mathbb{N} $ be arbitrary values. 
		Let $ {\mathscr{G}_c}_0 $ be a fixed arbitrary MAG
		satisfying Lemma~\ref{lemmaLabeledMAG} 
		such that, for every  $ i , j \leq p $, we have $ | \mathscr{A}( {\mathscr{G}_c}_0 )[i] | = | \mathscr{A}( {\mathscr{G}_c}_0 )[j] | = n_0 \in \mathbb{N}$. 
		Then, we build another arbitrary MAG $ {\mathscr{G}_c}_1 $ 
		such that, for every  $ i , j \leq p $, we have $ | \mathscr{A}( {\mathscr{G}_c}_1 )[i] | = | \mathscr{A}( {\mathscr{G}_c}_1 )[j] | = n_0 + 1 = n_1 \in \mathbb{N}$. 
		From an iteration of this process, we will obtain an infinite family $ F_{ \mathscr{G}_c } = \big\{ {\mathscr{G}_c}_0 , {\mathscr{G}_c}_1 , \dots , {\mathscr{G}_c}_i , \dots \big\} $ with $  \left| \mathbb{V}( {\mathscr{G}_c}_{ i } ) \right| = \left( n_0 + i \right)^p $, where no assumption was taken regarding the presence or absence of composite edges in their respective edge sets $ \mathscr{E} $, so that any $ {\mathscr{G}_c}_i \in F_{ \mathscr{G}_c } $ can be defined by any chosen composite edge set $ \mathscr{E}( {\mathscr{G}_c}_i ) $. 
		In addition, there is a total order with respect to the set of all composite vertices $ \mathbb{V} $ such that $ \mathbb{V}( {\mathscr{G}_c}_i ) \subsetneq \mathbb{V}( {\mathscr{G}_c}_{ i + 1 } ) $, where $ i \geq 0 $. 
		Thus, the next step is to construct a recursive ordering of composite edges for each one of these MAGs.  Like in Lemma~\ref{lemmaLabeledMAG}, we will construct a recursively ordered sequence of composite edges, which is independent of $ \mathscr{E} $. From this sequence, the algorithms that the strings $  \mathrm{p'}_1 $ and $  \mathrm{p'}_2 $ represent will become immediately defined. 
		To achieve this proof, we know that there is an algorithm that applies to $ {\mathscr{G}_c}_0 $ the ordering satisfying the proof of Lemma~\ref{lemmaLabeledMAG}.  Let $ \left( \mathbb{E}_c( {\mathscr{G}_c}_i  ) \right) $ denote an arbitrary sequence $ \left( e_1, e_2 , \dots, e_{ \left| \mathbb{E}_c( {\mathscr{G}_c}_i  ) \right| } \right) $ of all possible composite edges of MAG $ {\mathscr{G}_c}_i $ with $ i \geq 0 $. Then, one applies the iteration of: 
		\begin{itemize}
			\item If $ \left( \mathbb{E}_c( {\mathscr{G}_c}_k  ) \right) $, where $ k \geq 0 $, is a sequence of composite edges such that, for every  $ {\mathscr{G}_c}_i $ with $ 0 \leq i \leq k $,  $ \left( \mathbb{E}_c( {\mathscr{G}_c}_i  ) \right) $ is a prefix\footnote{ Also note that a sequence is always a prefix of itself. } of $ \left( \mathbb{E}_c( {\mathscr{G}_c}_k  ) \right) $, then:
			\begin{enumerate}[label=(\roman*)]
				\item apply to $ \mathbb{V}( {\mathscr{G}_c}_{ k + 1 } ) $ the recursive ordering $ <_{ \mathbb{E}_c } $ satisfying Lemma~\ref{lemmaLabeledMAG} ; \label{stepFirstrecursiveorderingiteration}
				
				\item concatenate after the last element of $ \left( \mathbb{E}_c( {\mathscr{G}_c}_k  ) \right) $ the elements of $ \mathbb{E}_c( {\mathscr{G}_c}_{ i + 1 } ) $ that were not already present in $ \left( \mathbb{E}_c( {\mathscr{G}_c}_k  ) \right) $, while preserving the order relation $ <_{ \mathbb{E}_c } $ previously applied to $ \mathbb{V}( {\mathscr{G}_c}_{ k + 1 } ) $ in Step~\ref{stepFirstrecursiveorderingiteration}.
			\end{enumerate}
		\end{itemize}
		Note that $ p, n_0 \in \mathbb{N} $ are fixed values. 
		Thus, let $  \mathrm{p'}_1 $ be a fixed string that represents on a prefix universal Turing machine the algorithm that, given $ \left< a_1 , \dots , a_p \right> $ and $ \left< b_1 , \dots , b_p \right> $ as inputs, builds the sequence of composite edges by the iteration described before, starting from $ \mathbb{E}_c \left( {\mathscr{G}_c}_0  \right)$, such that, before each $(k+1)$-th step of this iteration: 
		\begin{enumerate}[label=(\roman{*})]
			\item if, for every $ i $ with $ 1 \leq i \leq p $, one has $ a_i \leq n_0 + k   $ and $ b_i \leq n_0 + k  $, then: 
			
			\begin{enumerate}
				\item search for $ \left( \left( a_1 , \dots , a_p \right) , \left( b_1 , \dots , b_p \right) \right) $ or $ \left( \left( b_1 , \dots , b_p \right) , \left( a_1 , \dots , a_p \right) \right) $ in this sequence; 
				
				\item if one of these pairs is found, it returns the index value of the first one of these pairs found in this sequence.

			\end{enumerate}
			\item else, continue the iteration; 
			%
		\end{enumerate}
		Note that one of these pairs must be always eventually found, 
		since every $ a_i , b_i \in \mathbb{N} $.
		Therefore, $ \mathrm{p'}_1 $ never outputs $0$.
		An analogous algorithm defines $  \mathrm{p'}_2  $ (and $  \mathrm{p'}_2  $ only outputs $ \left< 0 \right> $ for $j=0$), but searching for the $j$-th element in the sequence and returning the respective pair of tuples instead.

		
	\end{proof}
\end{lemmaundersection}

Thus, the sequence of composite edges of MAGs in this family that has a smaller set of composite vertices is always a prefix of the sequence of composite edges of the one that has a larger set of composite vertices. Note that we have kept the order (i.e., number os aspects) of all MAGs in this family fixed with the purpose of avoiding some prefix indexing asymmetries due to dovetailing natural numbers inside the composite vertices for different values of $p$.
Regarding multidimensional network complexity analysis, another fruitful future research is to investigate to what extent Lemma~\ref{lemmaLabeledfamilyofMAG} can hold for families of MAGs with distinct order and/or distinct aspects.

If one is interested in investigating the algorithmic information or irreducible content measures of data representations of MAGs with the same order (i.e., number of aspects), Lemma~\ref{lemmaLabeledfamilyofMAG} guarantees that one can find a single way of encoding a MAG that applies at the same time to every possible configuration of presence or absence of composite edges.

Now, one of the immediate properties of a recursively labeled family of MAGs $ \mathscr{G}_c $ is that the algorithmic information contained in the composite edge set of such MAGs is tightly associated with the characteristic string in Lemma~\ref{lemmaBasicMAGandstrings}. 
To achieve such result, we will replace $ \left< \left< \tau( \mathscr{G}_c ) \right> , \mathrm{p}_2 \right> $ with $ \mathrm{p'}_2   $,  $ \left< \left< \tau( \mathscr{G}_c ) \right> , \mathrm{p}_1 \right> $ with $ \mathrm{p'}_1   $, and  $ \left< \left< \tau( \mathscr{G}_c ) \right> , p \right> $ with $ p  $ in the proofs of Lemma~~\ref{lemmaBasicMAGandstrings}.

\begin{corollaryafterlemmaundersection}\label{corFamilyoflabeledMAGandstrings}
	Let $  F_{ \mathscr{G}_c }   $ be a recursively labeled family (as in Definition~\ref{defLabeledfamilyofMAG}) of simple MAGs $ \mathscr{G}_c $.
	Then, for every $ \mathscr{G}_c  \in F_{ \mathscr{G}_c } $ and $ x \in \{ 0 , 1 \}^* $,
	where $ x $ is the characteristic string of $  \mathscr{G}_c  $, the following relations hold:
	
	\begin{align}
	\label{corFamilyoflabeledMAGandstrings1} C( \mathscr{E}( \mathscr{G}_c  ) \, | \, x  ) \leq K( \mathscr{E}( \mathscr{G}_c  ) \, | \, x  ) + \mathbf{O}(1)  & =   \mathbf{O}(1)  \\
	\label{corFamilyoflabeledMAGandstrings2} C( x \, | \, \mathscr{E}( \mathscr{G}_c  )  )  \leq K( x \, | \, \mathscr{E}( \mathscr{G}_c  )  ) + \mathbf{O}(1)  & =   \mathbf{O}(1) \\
	\label{corFamilyoflabeledMAGandstrings3} K( x ) & = K( \mathscr{E}( \mathscr{G}_c  ) )  \pm \mathbf{O}\left( 1 \right) \\
	\label{corFamilyoflabeledMAGandstrings4} I_A( x ; \mathscr{E}( \mathscr{G}_c  ) ) = I_A( \mathscr{E}( \mathscr{G}_c  ) ; x ) \pm \mathbf{O}(1) & = K( x ) - \mathbf{O}\left( 1 \right) 
	\end{align}
\end{corollaryafterlemmaundersection}

Regarding classical graphs, one can assume a constant $ p = | \mathscr{A}( \mathscr{G}_c ) | = 1 $ in in Lemma~\ref{lemmaLabeledfamilyofMAG}. 
Thus, since the composite edge sets $ \mathscr{E} $ were arbitrary, there will be a recursively labeled infinite family that contains all classical graphs $ G $. 
In other words, a classical graph is always a first-order MAG that belongs to a recursively labeled family that contains all MAGs of order $ p=1 $ and vertex labels in $ \mathbb{N} $. 
In this regard, from \cite{Abrahao2020cpublished,Abrahao2021publishednat} and the proof of Lemma~\ref{lemmaLabeledfamilyofMAG} with order $ p = 1 $, we will have that:

\begin{corollaryafterlemmaundersection}\label{corFamilygraphsandstrings}
	Let $ x \in \{ 0 , 1 \}^* $.
	Let $ G  $ be a classical graph from Definition~\ref{defClassicgraph}, where $ x $ is its characteristic string and $ \mathbb{E}_c( G ) = \{ \{ u , v \} \mid u , v \in V \} $.
	Then,
	\begin{align}
	\label{corFamilygraphsandstrings1} C( E( G  ) \, | \, x  ) \leq K( E( G  ) \, | \, x  ) + \mathbf{O}(1)  & =  \mathbf{O}( 1 )  \\
	\label{corFamilygraphsandstrings2} C( x \, | \, E( G  )  )  \leq K( x \, | \, E( G  )  ) + \mathbf{O}(1)  & =  \mathbf{O}( 1 ) \\
	\label{corFamilygraphsandstrings3} K( x ) & = K( E( G  ) )  \pm \mathbf{O}( 1 ) \\
	\label{corFamilygraphsandstrings4} I_A( x ; E( G  ) ) = I_A( E( G  ) ; x ) \pm \mathbf{O}(1) & = K( x ) - \mathbf{O}( 1 )
	\end{align}

\end{corollaryafterlemmaundersection}

Thus, these results ensure that one can apply an investigation of algorithmic randomness to finite MAGs analogously to classical graphs. 
In particular, for some families of MAGs in which the companion tuple does not add irreducible information in recursively ordering the composite edges---like those satisfying Definition~\ref{defLabeledfamilyofMAG}, 
Corollary~\ref{corFamilygraphsandstrings} ends up showing that our definitions and constructions of recursive labeling are consistent with the statements and constructions in \cite{Zenil2018a,Zenil2014,Li1997,Buhrman1999,Zenil2017a}. 
In the next section, we will show the existence of algorithmically random MAGs from a widely known example of $1$-random real number.

\section{A family of algorithmically random finite multiaspect graphs}\label{sectionK-randommultiaspectgraph}

One of the goals of this article is to show the existence of an infinite family of MAGs that contains a nesting sequence of MAGs in which one is a subMAG of the other. 
For this purpose, we will use an infinite $ \mathbf{O}(1) $-K-random (i.e., $1$-random) binary sequence as the source of information to build the edge set $ \mathscr{E} $. This is the main idea of our construction. 

We will give a constructive method for building an (finite) edge set $ \mathscr{E}( \mathscr{G}_c ) $ that is algorithmic-informationally equivalent to the $ n $ bits of $ \Omega $. 
Therefore, unlike the usage of C-random finite binary sequences like in Lemma \ref{lemmaFractionofrandomgraphs} from \cite{Li1997,Buhrman1999}, one can achieve a method for constructing a collection of \emph{prefix algorithmically random} MAGs (or graphs) using an infinite $ \mathbf{O}(1) $-K-random sequence as source. 


The key idea is to define a direct bijection between a recursively ordered sequence of composite edges, which in turn defines the composite edge sets $ \mathscr{E} ( \mathscr{G}_c ) $, and the bits of e.g. $ \Omega $. 
As an immediate consequence, $ \mathscr{E}( \mathscr{G}_c ) $ will be $ \mathbf{O}(1) $-K-random, that is, algorithmically random with respect to prefix algorithmic complexity (see Sections \ref{subsubsectionTMandAIT} and \ref{subsubsectionAIT}). 
Further, from previously established relations between K-randomness and C-randomness restated in Section \ref{subsubsectionAIT} and from Theorem \ref{thmMAGisomorphism}, we will show in Section~\ref{sectionC-randomMAGs} that this MAG is isomorphically equivalent to a $ \mathbf{O}( \lg( | \mathbb{V}( \mathscr{G}_c ) | ) ) $-C-random classical graph (see Section \ref{subsubsectionARGdefinitions}). Therefore, promptly enabling a direct application of the results in Section \ref{subsubsectionARG} to this MAG.


First, we define the prefix algorithmic randomness of composite edge set strings: 

\begin{definitionundersection}\label{defK-randomMAGs}
	Let $  F_{ \mathscr{G}_c }   $ be a family of simple MAGs whose composite edge set strings follow the same ordering.
	We say a simple MAG $ \mathscr{G}_c \in F_{ \mathscr{G}_c } $ (as in Definition~\ref{defSimplifiedMAG}) is (weakly)  $   \delta( \mathscr{G}_c ) $-K-random with respect to $  F_{ \mathscr{G}_c }   $ \textit{iff} it satisfies
	\[
	K( \mathscr{E}( \mathscr{G}_c ) ) \geq \binom{ \left| \mathbb{V}( \mathscr{G}_c ) \right| }{ 2 } -  \delta( \mathscr{G}_c )
	\]
\end{definitionundersection}

Thus, a $ \mathbf{O}(1) $-K-random MAG $ \mathscr{G}_c $ is an undirected MAG without self-loops with a network topology (which is in fact determined by the edge set $ \mathscr{E} $) that, for every composite edge set string in which the ordering of composite edges is unique for the entire family $  F_{ \mathscr{G}_c }   $, can only be compressed up to a constant that does not depend\footnote{ Although it can depend on the choice of the family or the universal programming language.} on the choice of $ \mathscr{G}_c $ in the family $  F_{ \mathscr{G}_c }   $. 
In other words, if one assumes $ \left< \mathscr{E}( \mathscr{G}_c ) \right> $ as the natural way to encode $ \mathscr{G}_c $, then one roughly needs the same number of bits of algorithmic information as the total number of possible composite edges in order to decide the existence or non existence of a composite edge. 
This follows the same intuition behind the definition of $ \mathbf{O}(1) $-K-random \emph{finite} sequences. 
Additionally, it bridges\footnote{ See Section~\ref{subsubsectionAIT}.} \emph{plain algorithmic randomness} (i.e., C-randomness) in classical graphs from \cite{Li1997,Buhrman1999} and \emph{prefix algorithmic randomness} (i.e., K-randomness) in multidimensional networks.
In this regard, Definition~\ref{defK-randomMAGs} differs from Definition~\ref{defC-randomgraph} in \cite{Li1997,Buhrman1999}, which takes into account the conditional plain algorithmic complexity of the edge set given the number of vertices. Nevertheless, the reader may notice that we have from Lemma~\ref{lemmaBasicAIT} that, for every $  \mathbf{O}(1) $-K-random MAG $ \mathscr{G}_c $, one also has that
\[
K( \mathscr{E}( \mathscr{G}_c )  \, \mid \, \left| \mathbb{V}( \mathscr{G}_c ) \right| ) \geq \binom{ \left| \mathbb{V}( \mathscr{G}_c ) \right| }{ 2 } - \mathbf{O}( \lg( \left| \mathbb{V}( \mathscr{G}_c ) \right| ) )
\text{ .}
\] 
\noindent That is, for $  \mathbf{O}(1) $-K-random MAGs $ \mathscr{G}_c $, informing the quantity of composite vertices to compress the composite edge set cannot save much more bits than the quantity necessary to compute this very informed quantity of composite vertices. 
Thus, one may define: 

\begin{definitionundersection}\label{defStronglyK-randomMAGs}
	Let $  F_{ \mathscr{G}_c }   $ be a family of simple MAGs whose composite edge set strings follow the same ordering.
	We say a simple MAG $ \mathscr{G}_c \in F_{ \mathscr{G}_c } $ with 
	\[
	K( \mathscr{E}( \mathscr{G}_c )  \, \mid \, \left| \mathbb{V}( \mathscr{G}_c ) \right| ) \geq \binom{ \left| \mathbb{V}( \mathscr{G}_c ) \right| }{ 2 } - \delta( \mathscr{G}_c ) 
	\]
	\noindent is a \emph{strongly}  $  \delta( \mathscr{G}_c ) $-K-random simple MAG $ \mathscr{G}_c $ with respect to $ F_{ \mathscr{G}_c } $. 
\end{definitionundersection}
This way, every \emph{weakly}  $  \mathbf{O}(1) $-K-random MAG $ \mathscr{G}_c $ is \emph{strongly}  $  \mathbf{O}( \lg( \left| \mathbb{V}( \mathscr{G}_c ) \right| ) ) $-K-random. In addition, it follows directly from Equation~\eqref{lemmaBasicAIT3} in Lemma~\ref{lemmaBasicAIT} that every $  \delta( \mathscr{G}_c ) $-C-random MAG $ \mathscr{G}_c $ (as we will define in Section~\ref{sectionC-randomMAGs}) is \emph{strongly}  $ \left( \delta( \mathscr{G}_c ) + \mathbf{O}(1) \right) $-K-random. However, the investigation of strongly $  \delta( \mathscr{G}_c ) $-K-random MAGs is not in the scope of this article and we will only deal with the weak case hereafter. This is the reason we have left the term ``weakly'' between parenthesis in Definition~\ref{defK-randomMAGs}, so that we will omit this term in this article.

It is also important to note that, 
despite the notation being similar, when we talk about $  \mathbf{O}(1) $-K-randomness of MAGs it refers to prefix algorithmic randomness of finite objects (i.e., objects with a representation in a finite binary sequence) and when we talk about $  \mathbf{O}(1) $-K-randomness of real numbers (as in Definition~\ref{defK-randomnessofreals}) it refers to prefix algorithmic randomness of an infinite binary sequence. However, as we will see in Section~\ref{subsectionNestedMAGs}, there will be a strict relation between initial segments of $  \mathbf{O}(1) $-K-random real numbers and nested subMAGs (or subgraphs) of a $  \mathbf{O}(1) $-K-random infinite nesting family of finite MAGs.
Thus, algorithmic randomness of individual (finite) MAGs (as in Definition~\ref{defK-randomMAGs}) should not be confused with the algorithmic randomness of nesting families of MAGs (as we will see in Definition~\ref{defK-randomnestedfamilyofMAGs}).
From AIT, it is important to remember that, although there are prefix algorithmically random infinite sequences (whose initial segments are incompressible up to a constant as in Theorem~\ref{thmOmeganumber}), the same does not hold for a plain algorithmic randomness of infinite sequences (i.e., infinite sequences $s$ with $ C( s \upharpoonright_n ) \geq n - \mathbf{O}(1) $). 
The reader may also want to see \cite{Li1997,Downey2010,Calude2002} for more properties and subtleties regarding algorithmically random \emph{finite} sequences and algorithmically random \emph{infinite} sequences.

Now, with the purpose of showing the existence of an individual $ \mathbf{O}(1) $-K-random MAG $ \mathscr{G}_c $, we will just combine previous results in algorithmic information theory with the ones that we have achieved in Section~\ref{sectionRecursivelylabeledMAGs}. 
Note that, although both results stem from the same construction in Lemma~\ref{lemmaLabeledfamilyofMAG}, the following lemma should not be confused with Lemma~\ref{lemmaNestedfamilyofMAGs} (or Theorem~\ref{thmNestedfamilyofK-randomMAGs}).

\begin{lemmaundersection}\label{lemmaK-randomMAGs}
	There is a recursively labeled \emph{infinite} family $ F_{ \mathscr{G}_c } $  (as in Definition~\ref{defLabeledfamilyofMAG}) of simple MAGs $ \mathscr{G}_c $ such that each one of them is $ \mathbf{O}(1) $-K-random with respect to  
	a recursively labeled infinite family $ F'_{ \mathscr{G}_c } $ of every possible simple MAG with the same arbitrary order and equal-size aspects.
	\begin{proof}
		From Lemma~\ref{lemmaLabeledfamilyofMAG}, we know there will be an infinite family $ F'_{ \mathscr{G}_c } $ that is recursively labeled with arbitrary presence or absence of composite edges in each MAG in this family. From Theorem~\ref{thmOmeganumber}, we have that
		\[
		K( \Omega \upharpoonright_n ) \geq n - \mathbf{O}( 1 )
		\]
		\noindent where $ n \in \mathbb{N} $ is arbitrary.
		Without loss of generality, since $ F'_{ \mathscr{G}_c } $ contains arbitrary arrangements of presence or absence of composite edges, we can now define family $ F_{ \mathscr{G}_c } $ as a subset of $ F'_{ \mathscr{G}_c } $ in which, for every  $ \mathscr{G}_c  \in F_{ \mathscr{G}_c } \subset F'_{ \mathscr{G}_c } $ with $ n = \left| \mathbb{E}_c( \mathscr{G}_c )   \right| $, we have that
		\[
		e_j  \in \mathscr{E}( \mathscr{G}_c  ) \iff \text{ the $j$-th digit in $ \Omega \upharpoonright_n $ is $1$}
		\text{ ,}
		\]
		\noindent where $ 1 \leq j \leq n \in \mathbb{N} $. 
		Then, for every $ \mathscr{G}_c  \in F_{ \mathscr{G}_c } $, $ \Omega \upharpoonright_n $ is a characteristic string.
		As a consequence, we will have from Corollary~\ref{corFamilyoflabeledMAGandstrings} that
		\[
		K( \mathscr{E}( \mathscr{G}_c  ) )  \pm \mathbf{O}\left( 1 \right) = K( \Omega \upharpoonright_n ) \geq n - \mathbf{O}( 1 )
		\text{ .}
		\] 
		\noindent  
		Therefore, since
		\[
		\binom{ \left| \mathbb{V}( \mathscr{G}_c ) \right| }{ 2 } = \frac{ { \left| \mathbb{V}( \mathscr{G}_c ) \right| }^2 - { \left| \mathbb{V}( \mathscr{G}_c ) \right| } }{ 2 } =  \left| \mathbb{E}_c( \mathscr{G}_c )   \right| = n
		\text{ ,}
		\]
		\noindent we will have that 
		\[
		K( \mathscr{E}( \mathscr{G}_c  ) )  \geq \binom{ \left| \mathbb{V}( \mathscr{G}_c ) \right| }{ 2 }  - \mathbf{O}( 1 )
		\] 
		
	\end{proof}
\end{lemmaundersection}

Additionally, since classical graphs are first-order MAGs $ \mathscr{G}_c $, the following corollary holds as a consequence of Corollary~\ref{corFamilygraphsandstrings}:

\begin{corollaryafterlemmaundersection}\label{corK-randomclassicgraphs}
	There is an infinite number of classical graphs $ G $ (as in Definition~\ref{defClassicgraph}) that are $ \mathbf{O}(1) $-K-random with respect to the recursively labeled family of all possible classical graphs.
	In particular, those in which their characteristic strings are long enough initial segments of a $\mathbf{O}(1)$-K-random real number.
\end{corollaryafterlemmaundersection}

\subsection{ An infinite family of nested multiaspect subgraphs}\label{subsectionNestedMAGs}

We know that a real number is $ \mathbf{O}(1) $-K-random if and only if it is weakly K-random for every initial segment (i.e., every prefix)---see Definition~\ref{defK-randomnessofreals}---of its representation in an infinite binary sequence \cite{Calude2002,Downey2010,Li1997}. 
Thus, asking the same about prefix algorithmically random data structured representations (in our case, strings) of MAGs or graphs would be a natural consequence of the previous results we have presented in this article. 
In fact, we will see that the same idea can be captured by nesting subgraphs of subgraphs and so on. 

In this section, we will extend the notion of subgraphs\footnote{ As in Definition~\ref{defSubgraphs}. } to MAGs. 
Then, we will see in Theorem~\ref{thmNestedfamilyofK-randomMAGs} that there is an infinite family of MAGs (or of classical graphs in Corollary~\ref{corNestedfamilyofK-randomgraphs}) that ``behave like'' initial segments of a $ \mathbf{O}(1) $-K-random real number.

First, the following definitions are just an extension of the common definitions of subgraphs, as in Definition~\ref{defSubgraphs}:

\begin{definitionafternotationundersubsection}\label{defSubMAGs}
	Let  $  \mathscr{G'} $ and $  \mathscr{G} $ be MAGs as in Definition~\ref{defMAG}.
	We say a MAG $  \mathscr{G'} $ is a \emph{multiaspect subgraph} (subMAG) of a MAG $  \mathscr{G} $, denoted as $ \mathscr{G'} \subseteq \mathscr{G} $, \textit{iff}
	\[
	\mathbb{V}\left( \mathscr{G'} \right) \subseteq \mathbb{V}\left( \mathscr{G} \right) 
	\; \land \;
	\mathscr{E}\left( \mathscr{G'} \right) \subseteq \mathscr{E}\left( \mathscr{G} \right) 
	\]
	
	\begin{definitionafternotationundersubsection}\label{defVertexinducedsubMAGs}
		We say a MAG $ \mathscr{G'} $ is a \emph{vertex-induced subMAG} of MAG $ \mathscr{G} $ \textit{iff} 
		\[
		\mathbb{V}\left( \mathscr{G'} \right) \subseteq \mathbb{V}\left( \mathscr{G} \right) 
		\] 
		\noindent and, for every $ \mathbf{ u } , \mathbf{ v } \in \mathbb{V}\left( \mathscr{G'} \right)  $,
		\[
		(  \mathbf{ u }  , \mathbf{ v } ) \in \mathscr{E}\left( \mathscr{G} \right)  \, \implies \, (  \mathbf{ u }  , \mathbf{ v } ) \in \mathscr{E}\left( \mathscr{G'} \right)
		\text{ .} 
		\]
		\noindent In addition, we denote this vertex-induced subMAG $ \mathscr{G'} $ as $ \mathscr{G} \left[ \mathbb{V}\left( \mathscr{G'} \right)  \right] $.
	\end{definitionafternotationundersubsection}
\end{definitionafternotationundersubsection}

Note that the composite vertex sets $ \mathbb{V} $'s in Definitions~\ref{defSubMAGs} and~\ref{defVertexinducedsubMAGs} could have a different number of aspects (i.e., node dimensions).
However, for the present purposes, one may take only the cases in which both MAGs $  \mathscr{G'} $ and $  \mathscr{G} $ have the same order. 

Now, we can construct a family of nested subMAGs in a way such that there is a total order for the subgraph operation. In this manner, for every two elements of this family, one of them must be a subMAG of the other. First, we will define a nesting family of MAGs in Definition~\ref{defNestedfamilyofMAGs}. Then, we will prove the existence of a nesting family that is recursively labeled and infinite in Lemma~\ref{lemmaNestedfamilyofMAGs}.

\begin{definitionafternotationundersubsection}\label{defNestedfamilyofMAGs}
	We say a family $  F_{ \mathscr{G} }^* $ of MAGs $ \mathscr{G} $ (as in Definition~\ref{defMAG}) is a \emph{nesting family} of MAGs $ \mathscr{G} $ \textit{iff}, for every $ \mathscr{G} , \mathscr{G'}, \mathscr{G''}  \in F_{ \mathscr{G} }^*  $, the following hold:
	\begin{align}
		\mathscr{G'} \subseteq \mathscr{G} \, 
		\land \, 
		\mathscr{G} \subseteq \mathscr{G'}
		&\implies
		\mathscr{G} = \mathscr{G'} \\
%
		\mathscr{G'} \subseteq \mathscr{G} \, 
		\land \, 
		\mathscr{G} \subseteq \mathscr{G''}
		&\implies
		\mathscr{G'} \subseteq \mathscr{G''} \\
%
		\mathscr{G'} \subseteq \mathscr{G} \; 
		&\lor \; 
		\mathscr{G} \subseteq \mathscr{G'}
	\end{align}
	
	\begin{definitionafternotationundersubsection}\label{defVertexinducednestedfamilyofMAGs}
		We say a family $  F_{ \mathscr{G} }^{v*} $ of MAGs $ \mathscr{G} $ (as in Definition~\ref{defMAG}) is a \emph{vertex-induced nesting family} of MAGs $ \mathscr{G} $ \textit{iff}, for every $ \mathscr{G} , \mathscr{G'}, \mathscr{G''}  \in F_{ \mathscr{G} }^{v*}  $, the following hold:
		\begin{align}
			\mathscr{G'} = \mathscr{G} \left[ \mathbb{V}\left( \mathscr{G'} \right)  \right] \subseteq \mathscr{G} \, 
			\land \, 
			\mathscr{G} = \mathscr{G'} \left[ \mathbb{V}\left( \mathscr{G} \right)  \right] \subseteq \mathscr{G'}
			&\implies
			\mathscr{G} = \mathscr{G'} \\
%
			\mathscr{G'} = \mathscr{G} \left[ \mathbb{V}\left( \mathscr{G'} \right)  \right] \subseteq \mathscr{G} \, 
			\land \, 
			\mathscr{G} = \mathscr{G''} \left[ \mathbb{V}\left( \mathscr{G} \right)  \right] \subseteq \mathscr{G''}
			&\implies
			\mathscr{G'} = \mathscr{G''} \left[ \mathbb{V}\left( \mathscr{G'} \right)  \right] \subseteq \mathscr{G''} \\
%
			\mathscr{G'} = \mathscr{G} \left[ \mathbb{V}\left( \mathscr{G'} \right)  \right] \subseteq \mathscr{G} \; 
			&\lor \; 
			\mathscr{G}= \mathscr{G'} \left[ \mathbb{V}\left( \mathscr{G} \right)  \right] \subseteq \mathscr{G'}  
		\end{align}
	\end{definitionafternotationundersubsection}
\end{definitionafternotationundersubsection}

It follows directly from these definitions that every vertex-induced nesting family in Definition~\ref{defVertexinducednestedfamilyofMAGs} is a nesting family in Definition~\ref{defNestedfamilyofMAGs}. 

As we will see in Lemma~\ref{lemmaNestedfamilyofMAGs}, one can define a vertex-induced nesting family of MAGs with order $p$ that is recursively labeled and infinite. 
Moreover, if one fix the recursive labeling for every possible composite edge set $ \mathscr{E} $ of these MAGs, there will be a non-denumerable amount of these families.
In turn, each one of these families belong to a larger recursively labeled family of all possible MAGs with the same order $p$.
The key idea is to bring the same recursive ordering of composite edges from Lemma~\ref{lemmaLabeledfamilyofMAG}. 

\begin{lemmaundersection}\label{lemmaNestedfamilyofMAGs}
	There is a non-denumerable amount of (vertex-induced) infinite nesting families $ F_{ \mathscr{G}_c }^{v*} $ of simple MAGs $ \mathscr{G}_c $, such that each one of these families is a subset of a recursively labeled infinite family that contains every possible MAG with the same order and the same set of composite vertices as any of the members of the families $ F_{ \mathscr{G}_c }^{v*} $.
	In particular, every real number $ x \in \left[ 0 , 1 \right] \subset \mathbb{R} $ with an infinite fractional part can univocally determine the presence or absence of a composite edge in each $ \mathscr{G}_c \in F_{ \mathscr{G}_c }^{v*} $. 
	
	\begin{proof}
		We will only prove the second part of the theorem, since we know that the cardinality of the set of real numbers $ x \in \left[ 0 , 1 \right] \subset \mathbb{R} $ with infinite fractional part in its binary representation is non-denumerable.
		Therefore, we will prove that an arbitrary real number $ x \in \left[ 0 , 1 \right] \subset \mathbb{R} $ with an infinite fractional part can univocally determine the presence or absence of a composite edge for every $ \mathscr{G}_c \in F_{ \mathscr{G}_c }^{v*} $, where $ F_{ \mathscr{G}_c }^{v*} $ is an infinite recursively labeled vertex-induced nesting family of MAGs. 
		To achieve this, let $ x \in \left[ 0 , 1 \right] \subset \mathbb{R} $ be an arbitrary real number with an infinite fractional part. Let $ F'_{ \mathscr{G}_c } $ be a family of MAGs defined in the proof of Lemma~\ref{lemmaLabeledfamilyofMAG}. Thus, there is $ p \in \mathbb{N} $ such that, for every $ \mathscr{G}_c \in F'_{ \mathscr{G}_c }$ and $ i , j \leq p $, we have that $ \mathscr{A}(\mathscr{G}_c )[ i ] \subset \mathbb{N} $, $ | \mathscr{A}( \mathscr{G}_c ) | = p $ and $ | \mathscr{A}( \mathscr{G}_c )[i] | = | \mathscr{A}( \mathscr{G}_c )[j] |$. Hence, by construction of $ F'_{ \mathscr{G}_c } $ in the proof of Lemma~\ref{lemmaLabeledfamilyofMAG}, for every $ \mathscr{G}_c ,  \mathscr{G'}_c , \mathscr{G''}_c \in F'_{ \mathscr{G}_c } $, we will have that: 
		\begin{align}
		\label{stepAntisymmetry} \mathbb{V}\left( \mathscr{G'}_c \right) \subseteq \mathbb{V}\left( \mathscr{G}_c \right) \, 
		\land \, 
		\mathbb{V}\left( \mathscr{G}_c \right) \subseteq \mathbb{V}\left( \mathscr{G'}_c \right)
		&\implies
		\mathbb{V}\left( \mathscr{G}_c \right) = \mathbb{V}\left( \mathscr{G'}_c \right) \, ; \\ 
		\mathbb{V}\left( \mathscr{G'}_c \right) \subseteq \mathbb{V}\left( \mathscr{G}_c \right) \, 
		\land \, 
		\mathbb{V}\left( \mathscr{G}_c \right) \subseteq \mathbb{V}\left( \mathscr{G''}_c \right)
		&\implies
		\mathbb{V}\left( \mathscr{G'}_c \right) \subseteq \mathbb{V}\left( \mathscr{G''}_c \right) \, ; \\
		\mathbb{V}\left( \mathscr{G'}_c \right) \subseteq \mathbb{V}\left( \mathscr{G}_c \right) \; 
		& \lor \; 
		\mathbb{V}\left( \mathscr{G}_c \right) \subseteq \mathbb{V}\left( \mathscr{G'}_c \right) \label{stepConnexproperty} \, ; \\
		\mathbb{V}\left( \mathscr{G}_c \right) \subseteq \mathbb{V}\left( \mathscr{G'}_c \right)
		& \implies 
		\left| \mathbb{E}_c( \mathscr{G}_c ) \right| 
		\leq
		\left| \mathbb{E}_c( \mathscr{G'}_c )  \right| \label{stepSmartorderingcondition} \, ;
		\end{align}
		\begin{align}
		\begin{aligned}\label{stepSmartordering}
			\text{ for every } e_i \in \mathbb{E}_c( \mathscr{G}_c ) \text{ and $ e'_j \in \mathbb{E}_c( \mathscr{G'}_c ) $} &\text{ with $ i \leq  \left| \mathbb{E}_c( \mathscr{G}_c ) \right|$ and $ j \leq \left| \mathbb{E}_c( \mathscr{G'}_c )  \right| $, } \\
			\mathbb{V}\left( \mathscr{G}_c \right) \subseteq \mathbb{V}\left( \mathscr{G'}_c \right) \; 
			\land \;
			j \leq \left| \mathbb{E}_c( \mathscr{G}_c )   \right| 
			&\implies 
			i = j
			\text{ .}
		\end{aligned}
		\end{align}
		\noindent Now, from Equations~\eqref{stepSmartorderingcondition} and~\eqref{stepSmartordering}, we construct a family  $ F_{ \mathscr{G}_c }^{v*} \subset F'_{ \mathscr{G}_c } $ as follows: 
		\begin{enumerate}[label= (\alph*)]
			\item 	if $ \mathbb{V}\left( \mathscr{G}_c \right) \subseteq \mathbb{V}\left( \mathscr{G'}_c \right) $, then 
			\begin{equation*}
			\begin{split}
			n &  \; \; \coloneq \binom{\left| \mathbb{V}\left( \mathscr{G}_c \right) \right| }{ 2 } 
			\leq 
			\binom{\left| \mathbb{V}\left( \mathscr{G'}_c \right) \right| }{ 2 } \\
			& \; \; \text{and} \\
			e_i  \in \mathscr{E}( \mathscr{G}_c  ) & \iff \text{ the $i$-th digit in $ x \upharpoonright_n $ is $1$ } \\
			& \; \; \text{and} \\
			e_j  \in \mathscr{E}( \mathscr{G'}_c  ) & \iff \text{ the $j$-th digit in $ x \upharpoonright_n $ is $1$ ;}
			\end{split}
			\end{equation*}
			
			\item  if $ \mathbb{V}\left( \mathscr{G'}_c \right) \subseteq \mathbb{V}\left( \mathscr{G}_c \right) $, then 
			\begin{equation*}
			\begin{split}
			n & \; \; \coloneq \binom{\left| \mathbb{V}\left( \mathscr{G'}_c \right) \right| }{ 2 }
			\leq
			\binom{\left| \mathbb{V}\left( \mathscr{G}_c \right) \right| }{ 2 } \\
			& \; \; \text{and} \\
			e_i  \in \mathscr{E}( \mathscr{G}_c  ) & \iff \text{ the $i$-th digit in $ x \upharpoonright_n $ is $1$ } \\
			& \; \; \text{and} \\
			e_j  \in \mathscr{E}( \mathscr{G'}_c  ) & \iff \text{ the $j$-th digit in $ x \upharpoonright_n $ is $1$ ;}
			\end{split}
			\end{equation*}
		\end{enumerate}

		\noindent To prove that this construction can always be correctly applied infinitely many often, note that, since $ F'_{ \mathscr{G}_c }  $ is infinite and Equations~\eqref{stepAntisymmetry} and~\eqref{stepConnexproperty} hold, we have that the mutually exclusive disjunction 
		\[
		\mathbb{V}\left( \mathscr{G'}_c \right) \subseteq \mathbb{V}\left( \mathscr{G}_c \right) \; 
		\lor \; 
		\mathbb{V}\left( \mathscr{G}_c \right) \subseteq \mathbb{V}\left( \mathscr{G'}_c \right)
		\]
		holds infinitely many often in $ F'_{ \mathscr{G}_c } $. 
		Finally, to prove that each real number $x$ generates a unique infinite nesting family, except for MAG automorphisms, note that the ordering of composite edges are fixed for the entire family $ F'_{ \mathscr{G}_c } $ from which each infinite nesting family is a subset.
		Therefore, any reordering (or re-indexing) of composites edges
		(including those by composite vertex re-labeling) 
		leads to a MAG that does not belong to $ F'_{ \mathscr{G}_c } $.
		In addition, any MAG automorphism applied to these infinite families that still follows the same recursively labeling of $ F'_{ \mathscr{G}_c } $ produces two equal characteristic strings.
	\end{proof}
\end{lemmaundersection}

In this way, a family of MAGs that satisfies Lemma~\ref{lemmaNestedfamilyofMAGs} immediately gives us an infinite sequence of nested subMAGs.
We have shown that each real number has a correspondent family of nested subMAGs such that any initial segment of this real number is a characteristic string of the respective subMAG. 
Note that, although the family $ F'_{ \mathscr{G}_c } $ and the infinite nesting families are composed of only finite objects and, therefore, are countably infinite sets, the non-denumerable set of infinite nesting families is a collection of subsets of $ F'_{ \mathscr{G}_c } $.  
The issue we are going to tackle now is whether there is a nesting chain of subMAGs that could behave like initial segments of a $1$-random infinite binary sequence.

\begin{definitionafternotationundersubsection}\label{defK-randomnestedfamilyofMAGs}
	We say an infinite nesting family $ F_{ \mathscr{G}_c }^{*} $ (as in Definition~\ref{defNestedfamilyofMAGs}) of simple MAGs $ \mathscr{G}_c $ (as in Definition~\ref{defSimplifiedMAG}) is $  \mathbf{O}(1) $-K-random \textit{iff}: 
	
	\begin{itemize}
		\item there is a recursively labeled infinite family $ F_{ \mathscr{G}_c } $ such that $ F_{ \mathscr{G}_c }^{*} \subseteq F_{ \mathscr{G}_c } $ and $ F_{ \mathscr{G}_c } $ contains every possible MAG with the same order and the same set of composite vertices as any of the members of $ F_{ \mathscr{G}_c }^{*} $;
		
		\item every $  \mathscr{G}_c \in  F_{ \mathscr{G}_c }^{*} $ is $  \mathbf{O}(1) $-K-random (as in Definition~\ref{defK-randomMAGs}) with respect to $ F_{ \mathscr{G}_c } $;
	\end{itemize}

\end{definitionafternotationundersubsection}

Thus, Definition~\ref{defK-randomnestedfamilyofMAGs} embeds into such a concept of prefix algorithmically random infinite nesting family the property of this family being a subset of a larger recursively labeled infinite family, which also contains any other alternative composite edge set $ \mathscr{E} $ with respect to the members of the former family.
This formalization of prefix algorithmic randomness for infinite nesting families is relative in the sense that it depends on the existence of a larger encompassing family whose distinct feature (or structural characteristic) is being recursively labeled.
The usual notion of a family of graphs being understood as a set of graphs that share a common distinctive structural characteristic---in our present case, being recursively labeled---is the reason we are investigating algorithmic randomness of infinite families of MAGs and not algorithmic randomness of infinite MAGs in general.
Indeed, this was one of the properties employed e.g. in the proof of Lemma~\ref{lemmaNestedfamilyofMAGs}.
Although the recursive labeling of $ F_{ \mathscr{G}_c } $ is unique and independent of the choice of any member of $ F_{ \mathscr{G}_c } $ and holds for every possible network topology (i.e, every composite edge set $ \mathscr{E} $), the implicit ordering of composite edges is fixed---which is an expected property to hold for the present purposes of investigating data representations of multidimensional networks.
On the other hand, beyond data representation of families of MAGs that are recursively labeled, we suggest future research in order to study algorithmic randomness of infinite MAGs in general without the property of fixing the ordering (or indexing) of each composite edge, such as algorithmic randomness of structures in \cite{Khoussainov2014,Harrison-Trainor2019}. 

In a recursive labeling agnostic manner, one can e.g. define nesting families of MAGs whose composite edge set strings are computationally retrieved from initial segments of real numbers (and vice-versa):

\begin{definitionafternotationundersubsection}\label{defNestedfamilyofMAGsbyrealnumbers}
	Let $ x \in \left[ 0 , 1 \right] \subset \mathbb{R} $ be an arbitrary real number with an infinite fractional part. 
	We denote by $ F_{x} $ the nesting family $ F_{ \mathscr{G}_c }^{*} $ (as in Definition~\ref{defNestedfamilyofMAGs}) of simple MAGs $ \mathscr{G}_c $ (as in Definition~\ref{defSimplifiedMAG}) in which, for every $ \mathscr{G}_c \in F_{ \mathscr{G}_c }^{*} $ with $ n =  \left| \mathbb{E}_c( \mathscr{G}_c )   \right| $,
	\[
	K( \mathscr{E}( \mathscr{G}_c ) ) = K( x \upharpoonright_n ) \pm \mathbf{O}(1)
	\]
\end{definitionafternotationundersubsection}

In fact, we will see now that both Definitions~\ref{defK-randomnestedfamilyofMAGs} and~\ref{defNestedfamilyofMAGsbyrealnumbers} can be satisfied by a family that was already constructed for the proof of Lemma~\ref{lemmaK-randomMAGs}; and, thus, the following Theorem~\ref{thmNestedfamilyofK-randomMAGs} may be seen as particular case of Lemma~\ref{lemmaNestedfamilyofMAGs}.

\begin{thmafterlemmaundersection}\label{thmNestedfamilyofK-randomMAGs}
	There is a recursively labeled (vertex-induced) infinite nesting family $ F_{ \mathscr{G}_c }^{v*} $ (as in Definition~\ref{defVertexinducednestedfamilyofMAGs}) of simple MAGs $ \mathscr{G}_c $ that is $  \mathbf{O}(1) $-K-random. In particular, there is a $  \mathbf{O}(1) $-K-random recursively labeled (vertex-induced) infinite nesting family $ F_{ \Omega } $ (as in Definition~\ref{defNestedfamilyofMAGsbyrealnumbers}) of simple MAGs $ \mathscr{G}_c $.
	
	\begin{proof}
		Let $ F_{ \mathscr{G}_c }^{v*} $ be a family constructed as in the proof of Lemma~\ref{lemmaNestedfamilyofMAGs}. 
		Since the real number $ x \in \left[ 0 , 1 \right] \subset \mathbb{R} $ was arbitrary, we can assume e.g. $ x = \Omega $ (see Definition~\ref{defOmeganumber}) in this construction. 
		Therefore, we have from the proof of Lemma~\ref{lemmaNestedfamilyofMAGs} that this family immediately satisfies Lemma~\ref{lemmaK-randomMAGs} and every $  \mathscr{G}_c \in F_{ \mathscr{G}_c }^{v*} $ is $  \mathbf{O}(1) $-K-random with respect to the family satisfying Lemma~\ref{lemmaLabeledfamilyofMAG}. 
		In addition, since this family is a subset of the family satisfying Lemma~\ref{lemmaLabeledfamilyofMAG}, we will have from Corollary~\ref{corFamilyoflabeledMAGandstrings} that
		\[
		K( \mathscr{E}( \mathscr{G}_c ) ) = K( \Omega \upharpoonright_n ) \pm \mathbf{O}(1)
		\]
	\end{proof}
\end{thmafterlemmaundersection}

The reader is invited to note that the recursively labeling from Definition~\ref{defLabeledfamilyofMAG} that underlies the proofs of Lemmas~\ref{lemmaLabeledfamilyofMAG}, ~\ref{lemmaK-randomMAGs} and~\ref{lemmaNestedfamilyofMAGs} in fact guarantees a stronger equivalence into Theorem~\ref{thmNestedfamilyofK-randomMAGs} than the one in Definition ~\ref{defNestedfamilyofMAGsbyrealnumbers}.
As we have constructed,
the characteristic string $x$ is Turing equivalent to $ \left< \mathscr{E}( \mathscr{G}_c ) \right> $.
Of course, from Corollary~\ref{corFamilyoflabeledMAGandstrings}, this implies
\[
K( \mathscr{E}( \mathscr{G}_c ) ) = K( x \upharpoonright_n ) \pm \mathbf{O}(1)
\text{ ,}
\] 
but the inverse direction may be not true, as in a similar phenomenon to K-trivial infinite sequences \cite{Downey2010}.
Thus, an interesting future research is whether there is infinite nesting family $ F_y $ such that $ y \in \mathbb{R} $ is $1$-random, but $ F_y $ cannot be recursively labeled. 
That is, a family that satisfies Definition~\ref{defNestedfamilyofMAGsbyrealnumbers}, but cannot satisfy Definition~\ref{defK-randomnestedfamilyofMAGs}. 

In fact, as previously shown for classical graphs in \cite{Buhrman1999} and also referenced in \cite{Khoussainov2014}, we will later on show in Corollary~\ref{corK-randomnestedfamilyproperties} that the incompressibility of infinite nesting families of finite MAGs is robust to automorphisms beyond the one given by the identity isomorphism.
Moreover, it is easy to show that, in the same way a prefix algorithmically random infinite sequence keeps being algorithmically random\footnote{ Except for a constant that depends on the recursive functions performing the reordering of terms.} for every recursively bijective reordering\footnote{ That is, for every total computable function that performs a re-indexing of the elements of the sequence such that the inverse of this function is also computable and total.} of the sequence's elements, a prefix algorithmically random infinite nesting family (see Definition~\ref{defK-randomnestedfamilyofMAGs}) will keep to have prefix algorithmically random MAGs (as we will define in Section~\ref{sectionK-randommultiaspectgraph}) for every recursively bijective reordering of the infinite sequence of composite edges embedded in Definition~\ref{defLabeledfamilyofMAG}, should the latter be satisfied by a family e.g. from Lemma~\ref{lemmaLabeledfamilyofMAG}.
While the present work is concerned with algorithmic randomness for data representation (in particular, in the form of strings) of multidimensional objects, we also suggest future research on how the algorithmic randomness of infinite nesting families from Definition~\ref{defK-randomnestedfamilyofMAGs} relates to the algorithmic randomness of model-theoretic infinite structures as in \cite{Khoussainov2014,Harrison-Trainor2019}.

By using Corollary~\ref{corFamilygraphsandstrings} instead of Corollary~\ref{corFamilyoflabeledMAGandstrings}, and Corollary~\ref{corK-randomclassicgraphs} instead of Lemma~\ref{lemmaK-randomMAGs}, 
the following corollary can be achieved from the proof of Theorem~\ref{thmNestedfamilyofK-randomMAGs}:

\begin{corollaryafterlemmaundersection}\label{corNestedfamilyofK-randomgraphs}
	There is a recursively labeled (vertex-induced) infinite nesting family $ F_{ G }^{v*} $ (as in Definition~\ref{defVertexinducednestedfamilyofMAGs}) of classical graphs (as in Definition~\ref{defClassicgraph}) that is $  \mathbf{O}(1) $-K-random. In particular, there is a $  \mathbf{O}(1) $-K-random recursively labeled (vertex-induced) infinite nesting family $ F_{ \Omega } $ (as in Definition~\ref{defNestedfamilyofMAGsbyrealnumbers}) of classical graphs $ G $ (as in Definition~\ref{defSimplifiedMAG}). 
\end{corollaryafterlemmaundersection}

\section{Plain and prefix algorithmic randomness of finite multiaspect graphs}\label{sectionC-randomMAGs}

We have shown that algorithmic randomness regarding prefix algorithmic complexity, (i.e., prefix algorithmic randomness or K-randomness) in multiaspect graphs (MAGs) defines a class of MAGs with a topology that can only be described by the same amount of algorithmic information (except for a constant) as the number of all virtually possible connections. 
In Section~\ref{sectionK-randommultiaspectgraph}, these results were achieved by extending the concept of algorithmic randomness of classical graphs regarding plain algorithmic complexity (i.e., plain algorithmic randomness or C-randomness) in \cite{Li1997,Buhrman1999} to finite or infinite families of MAGs.
Therefore, a natural consequence would be studying the relation between (weakly) $ \mathbf{O}(1) $-K-random MAGs and $ \delta( \left| \mathbb{V}( \mathscr{G}_c ) \right| ) $-C-random MAGs.

One of the results in algorithmic information theory (see Theorem~\ref{thmK-randomandC-random}) is that one can retrieve a lower bound for the plain algorithmic complexity of finite initial segments of infinite binary sequences that are $ \mathbf{O}(1) $-K-random. Thus, in this section, we apply this same property to MAGs. In particular, we study  $ \delta(n) $-C-randomness in MAGs that are $ \mathbf{O}(1) $-K-random. 

\begin{definitionundersection}\label{defC-randomMAG}
	Let $  F_{ \mathscr{G}_c }   $ be a family of simple MAGs whose composite edge set strings follow the same ordering.
	We say a simple MAG $ \mathscr{G}_c \in F_{ \mathscr{G}_c } $ (as in Definition~\ref{defSimplifiedMAG}) is $ \delta( \left| \mathbb{V}( \mathscr{G}_c ) \right| ) $-C-random with respect to $ F_{ \mathscr{G}_c } $ \textit{iff} it satisfies
	\[
	C\left( \mathscr{E}( \mathscr{G}_c ) \, \big\mid \left| \mathbb{V}( \mathscr{G}_c ) \right| \right) 
	\geq 
	\binom{ \left| \mathbb{V}( \mathscr{G}_c ) \right| }{ 2 } - \delta( \left| \mathbb{V}( \mathscr{G}_c ) \right| )
	\]
	\noindent where 
	\[ \myfunc{ \delta }{ \mathbb{N} }{ \mathbb{N} }{ n  }{ \delta(n) } \] 
	\noindent is the randomness deficiency function.
\end{definitionundersection}

This definition directly extends Definition~\ref{defC-randomgraph} to MAGs, taking into account that Corollary~\ref{corClassicMAGisomorphism} (and, as we will show, Theorem~\ref{thmC-randomMAGsandgraphs}) gives us an isomorphic representation of a simple labeled MAG $ \mathscr{G}_c $ as a classical graph. Therefore, it enables a proper interpretation of previous results in \cite{Zenil2018a,Zenil2014,Li1997,Buhrman1999,Zenil2017a} into the context of MAGs. 

However, before studying some properties of $ \delta( \left| \mathbb{V}( \mathscr{G}_c ) \right| ) $-C-random MAGs, we will investigate the relation between $ \mathbf{O}(1) $-K-random MAGs and $ \delta( \left| \mathbb{V}( \mathscr{G}_c ) \right| ) $-C-random MAGs. The main idea is to construct MAGs from composite edge sets determined by binary strings that are prefixes of $ \mathbf{O}(1) $-K-random real numbers. This will give rise not only to $ \mathbf{O}(1) $-K-random MAGs, which are basically weakly K-random strings (see Definition~\ref{defWeakK-randomnessofstrings}), but also to $ \delta( \left| \mathbb{V}( \mathscr{G}_c ) \right| ) $-C-random MAGs.
Therefore, together with previous studies on algorithmic randomness, as restated in Theorem~\ref{thmK-randomandC-random}, we will now be able to obtain the following theorem:

\begin{thmafterlemmaundersection}\label{thmK-randomandC-randomMAGs}
	Let $  F_{ \mathscr{G}_c }   $ be a recursively labeled infinite family of simple MAGs $ \mathscr{G}_c $ (as in Definition~\ref{defLabeledfamilyofMAG}) such that, for every $ \mathscr{G}_c  \in F_{ \mathscr{G}_c }  $ and $ n \in \mathbb{N} $, if $ x \upharpoonright_n $ is its characteristic string and $ n = \left| \mathbb{E}_c( \mathscr{G}_c )   \right| $,
	then  $ x \in \left[ 0 , 1 \right] \subset \mathbb{R} $ is $ \mathbf{O}(1) $-K-random (as in Definition~\ref{defK-randomnessofreals}). 
	Therefore, every  MAG $ \mathscr{G}_c  \in F_{ \mathscr{G}_c } $ is $ \mathbf{O}( \lg( \left| \mathbb{V}( \mathscr{G}_c ) \right| ) ) $-C-random and (weakly) $ \mathbf{O}(1) $-K-random with respect to $ F_{ \mathscr{G}_c } $.
	In addition, there is such family $  F_{ \mathscr{G}_c }   $ with $ x =  \Omega  \in \left[ 0 , 1 \right] \subset \mathbb{R}  $.

	\begin{proof}
		First, we prove that every MAG $ \mathscr{G}_c  \in F_{ \mathscr{G}_c } $ is $ \mathbf{O}(1) $-K-random with respect to $ F_{ \mathscr{G}_c } $. 
		We have that, for every $ \mathscr{G}_c $ there is $ x \upharpoonright_n \, \in \{ 0 , 1 \}^* $ with $ n = l( x \upharpoonright_n ) = \left| \mathbb{E}_c( \mathscr{G}_c )   \right| $ and
		\[
		e  \in \mathscr{E}( \mathscr{G}_c  ) \iff \text{ the $j$-th digit in $ x \upharpoonright_n $ is $1$}
		\text{ ,}
		\]
		\noindent  where $ 1 \leq j \leq l( x \upharpoonright_n ) $, $ n \in \mathbb{N} $ and $ e \in \mathbb{E}_c( \mathscr{G}_c ) $. Hence, by hypothesis, we have that $ x \in \left[ 0 , 1 \right] \subset \mathbb{R} $ is $ \mathbf{O}(1) $-K-random.
		Thus, from Definition~\ref{defK-randomnessofreals} and Corollary~\ref{corFamilyoflabeledMAGandstrings}, we will have that 
		\begin{equation*}
		K( \mathscr{E}( \mathscr{G}_c  ) )  \pm \mathbf{O}\left( 1 \right) = K( x \upharpoonright_n ) \geq l( x \upharpoonright_n ) - \mathbf{O}(1) = \binom{ \left| \mathbb{V}( \mathscr{G}_c ) \right| }{ 2 } - \mathbf{O}(1)
		\end{equation*}
		\noindent Thus, from Definition~\ref{defK-randomMAGs}, we will have that every MAG $ \mathscr{G}_c  \in F_{ \mathscr{G}_c } $ is $ \mathbf{O}(1) $-K-random with respect to $ F_{ \mathscr{G}_c } $.
		Now, in order to prove that every MAG $ \mathscr{G}_c  \in F_{ \mathscr{G}_c } $ is $ \mathbf{O}( \lg( \left| \mathbb{V}( \mathscr{G}_c ) \right| ) ) $-C-random, note that Theorem~\ref{thmK-randomandC-random} implies that, if $ x \in \left[ 0 , 1 \right] \subset \mathbb{R} $ is $ \mathbf{O}(1) $-K-random, then 
		\begin{equation}\label{stepAITtrick}
		C( x \upharpoonright_n ) \geq n - K( n ) - \mathbf{O}(1) 
		\end{equation}
		Then, from Lemma~\ref{lemmaBasicAIT} and Corollary~\ref{corFamilyoflabeledMAGandstrings},
		\noindent we will have that
		\begin{align*}
		& \mathbf{O}( \lg( \left| \mathbb{V}( \mathscr{G}_c ) \right| ) ) + C( \mathscr{E}( \mathscr{G}_c  ) \, \big\mid \; \left| \mathbb{V}( \mathscr{G}_c ) \right|  ) + \mathbf{O}( \lg( \left| \mathbb{V}( \mathscr{G}_c ) \right| ) ) \pm \mathbf{O}(1) \geq \\
		& \geq n - \mathbf{O}( \lg( \left| \mathbb{V}( \mathscr{G}_c ) \right| ) ) - \mathbf{O}(1) = \binom{ \left| \mathbb{V}( \mathscr{G}_c ) \right| }{ 2 } - \mathbf{O}( \lg( \left| \mathbb{V}( \mathscr{G}_c ) \right| ) ) 
		\end{align*}
		Let $ \delta( \left| \mathbb{V}( \mathscr{G}_c ) \right| ) = \mathbf{O}( \lg( \left| \mathbb{V}( \mathscr{G}_c ) \right| ) ) $.
		Thus, from Definition~\ref{defC-randomMAG}, we have that $ \mathscr{G}_c $ is $ \mathbf{O}( \lg( \left| \mathbb{V}( \mathscr{G}_c ) \right| ) ) $-C-random with respect to $ F_{ \mathscr{G}_c } $. 
		In order to prove that there is such family $  F_{ \mathscr{G}_c }   $ with $ \Omega = x \in \left[ 0 , 1 \right] \subset \mathbb{R}  $, just use the one from the proof of Lemma~\ref{lemmaK-randomMAGs}.
	\end{proof}
\end{thmafterlemmaundersection}

With this result, we can study plain algorithmic randomness in $ \mathbf{O}(1) $-K-random infinite nesting families of MAGs. Thus, by choosing a family of MAGs that satisfies Theorem~\ref{thmNestedfamilyofK-randomMAGs} we will have from Corollary~\ref{corFamilyoflabeledMAGandstrings} that the conditions of Theorem~\ref{thmK-randomandC-randomMAGs} are immediately satisfied. 
Hence:

\begin{corollaryafterlemmaundersection}\label{thmK-randomandC-randomnestedsubMAGs}
	Let $  F_{ \mathscr{G}_c }^{v*}   $ be a recursively labeled $ \mathbf{O}(1) $-K-random infinite (vertex-induced) nesting family (as in Theorem~\ref{thmNestedfamilyofK-randomMAGs}) of simple MAGs $ \mathscr{G}_c $ (as in Definition~\ref{defLabeledfamilyofMAG}). Then,  every  MAG $ \mathscr{G}_c  \in F_{ \mathscr{G}_c }^{v*} $ is $ \mathbf{O}( \lg( \left| \mathbb{V}( \mathscr{G}_c ) \right| ) ) $-C-random with respect to $  F_{ \mathscr{G}_c }^{v*}   $. 
\end{corollaryafterlemmaundersection}

Furthermore, the same case for classical graphs applies as a particular case by employing Corollary~\ref{corNestedfamilyofK-randomgraphs} instead of Theorem~\ref{thmNestedfamilyofK-randomMAGs}:

\begin{corollaryafterlemmaundersection}
	Let $  F_{ G }^{v*}   $ be a recursively labeled (vertex-induced) infinite nesting $ \mathbf{O}(1) $-K-random family (as in Corollary~\ref{corNestedfamilyofK-randomgraphs}) of classical graphs $ G $ (as in Definition~\ref{defClassicgraph}). Then,  every  classical graph $ G \in F_{ G }^{v*} $ is $ \mathbf{O}( \lg( \left| V( G ) \right| ) ) $-C-random with respect to $  F_{ G }^{v*}   $. 
\end{corollaryafterlemmaundersection}

\subsection{Algorithmic information of the isomorphism between multiaspect graphs and graphs}\label{subsectionMAGgraphsisomorphism}

One may be interested in comparing the network topological information of a MAG with the network topological information of a graph.
Indeed, we will show that this is possible under a minor correction in the randomness deficiency, resulting from an algorithmic-informational cost of performing a transformation of a MAG into its isomorphically correspondent graph, and vice-versa.
Although such an isomorphism (as in Definition~\ref{defMAGandgraphsisomorphisms}) is an abstract mathematical equivalence, when dealing with structured data representations of objects (e.g., as the strings $ \left< \mathscr{E}( \mathscr{G}_c ) \right> $ or $ \left< E( G ) \right> $), this equivalence must always be performed by computable procedures. 
Therefore, there is always an associated necessary algorithmic information, which corresponds to the length of the minimum program that performs these transformations.
Theorem~\ref{thmC-randomMAGsandgraphs} below gives a worst-case algorithmic-informational cost in a logarithmic order of the network size for plain algorithmic complexity (i.e., for non-prefix-free universal programming languages in general).
In addition, in the prefix case (as in Definition~\ref{BdefK}), Theorem~\ref{BthmKrandomMAGsandgraphs} will show that there is an algorithmic-informational cost upper bounded by a positive constant that only depends on the chosen universal prefix-free programming language or on the choice of the pairing function $ \left< \cdot , \cdot \right> $.
The key idea of the following results derives directly from applying the equivalence of MAGs and graphs from Section~\ref{subsubsectionMultiaspect-graphs} and the same recursively labeling of families $  F_{ \mathscr{G}_c } $'s in order to achieve a correspondent recursively labeled family of classical graphs.

\begin{thmafterlemmaundersection}\label{thmC-randomMAGsandgraphs}
	Let $  F_{ \mathscr{G}_c } \neq \emptyset  $ be an arbitrary recursively labeled family of simple MAGs $ \mathscr{G}_c $ (as in Definition~\ref{defLabeledfamilyofMAG}).
	Then, for every $  \mathscr{G}_c  \in  F_{ \mathscr{G}_c } $,
	\begin{center}
		$ \mathscr{G}_c$ is $ \left( \delta( \left| \mathbb{V}( \mathscr{G}_c ) \right| ) + \mathbf{O}( \lg( \left| \mathbb{V}( \mathscr{G}_c ) \right|) ) \right) $-C-random \\
		$ \text{iff} $ \\
		$ G $ is $ \left( \delta( \left| V( G ) \right| ) + \mathbf{O}( \lg( \left| V( G ) \right| ) ) \right) $-C-random,
	\end{center}
	\noindent where $ G $ is isomorphic (as in Corollary~\ref{corClassicMAGisomorphism}) to $ \mathscr{G}_c $.
	
	\begin{proof}
		The existence and uniqueness of $ G $ is guaranteed by Corollary~\ref{corClassicMAGisomorphism}, which follows from the proof of Theorem~\ref{thmMAGisomorphism} in \cite{Wehmuth2016b} with a symmetric adjacency matrix. Thus, we will first describe a recursive procedure for constructing this unique isomorphic classical graph $ G $ from $ \mathscr{G}_c \in F_{ \mathscr{G}_c } $ and vice-versa. Then, it will only remain to prove that (except for the information necessary to compute the size of the graph): 
		\begin{center}
			$ \mathscr{G}_c$ is $ \delta( \left| \mathbb{V}( \mathscr{G}_c ) \right| ) $-C-random \textit{iff} $ G $ is $ \delta( \left| V( G ) \right| ) $-C-random 
		\end{center}
		\noindent In order to construct such classical graph $ G $ from $ \mathscr{G}_c \in F_{ \mathscr{G}_c } $, it is important to remember the proof of Theorem~\ref{thmMAGisomorphism} in \cite{Wehmuth2016b}. Assume here the same procedure described there for the existence of $ G $. Since $  \mathscr{G}_c $ belongs to a recursively labeled family, as in Definition~\ref{defLabeledfamilyofMAG}, then we can take the recursive bijective pairing function $ \left< \cdot , \cdot , \dots , \cdot \right> $ on which this recursive labeling holds for this family. Hence, since the recursive bijective pairing function $ \left< \cdot , \cdot , \dots , \cdot \right> $ is now fixed, there is a recursively bijective function 
		\[ 
		\myfunc{f}{  \mathbb{V}( \mathscr{G}_c )  }{ V( G ) = \{ 1 , \dots , n \}\subset \mathbb{N} }{ \left( a_1 , \dots , a_p \right) }{ f(\left( a_1 , \dots , a_p \right)   )  \in \mathbb{N} }
		\text{ ,} 
		\]
		\noindent where $ n = \left| V( G ) \right| = \left| \mathbb{V}( \mathscr{G}_c ) \right|  \in \mathbb{N} $, that performs a bijective relabeling between vertices of $ G $ and composite vertices of $ \mathscr{G}_c $. 
		Therefore, given $ \mathscr{E}( \mathscr{G}_c ) $ as input, there is an algorithm that reads the string $ \left< \mathscr{E}( \mathscr{G}_c ) \right> $ and replace each composite vertex by its corresponding label in $  V( G )  $ using function $f$, and then returns $ \left< E( G ) \right> $. 
		On the other hand, given $  E( G ) $ as input, there is an algorithm that reads this string $ \left< E( G ) \right> $ and replace each vertex by its corresponding label in $ \mathbb{V}( \mathscr{G}_c )  $ using function $f^{-1}$, and then returns $ \left< \mathscr{E}( \mathscr{G}_c ) \right> $. Thus, since $ \left| V( G ) \right| = \left| \mathbb{V}( \mathscr{G}_c ) \right|  \in \mathbb{N} $, we will have that
		\begin{equation}\label{equationWeaklyKrandom}
		\begin{aligned}
		K\left( \mathscr{E}( \mathscr{G}_c ) \right) 
		=
		K\left( E( G ) \right)  \pm \mathbf{O}(1)
		\end{aligned}
		\end{equation}
		and
		\begin{equation}\label{equationStronglyKrandom}
		\begin{aligned}
		K\left( \mathscr{E}( \mathscr{G}_c ) \, \big\mid \left| \mathbb{V}( \mathscr{G}_c ) \right| \right) 
		=
		K\left( E( G ) \, \big\mid \left| V( G ) \right| \right)  \pm \mathbf{O}(1)
		\end{aligned}
		\end{equation}
		Now, we split the proof in two cases: 
		\begin{enumerate}
			\item first, when 
			\[ K\left( \mathscr{E}( \mathscr{G}_c ) \, \big\mid \left| \mathbb{V}( \mathscr{G}_c ) \right| \right) 
			\leq
			K\left( E( G ) \, \big\mid \left| V( G ) \right| \right)  + \mathbf{O}(1) \text{ ;}\] 
			
			\item second, when 
			\[ K\left( \mathscr{E}( \mathscr{G}_c ) \, \big\mid \left| \mathbb{V}( \mathscr{G}_c ) \right| \right) + \mathbf{O}(1)
			\geq
			K\left( E( G ) \, \big\mid \left| V( G ) \right| \right)   \text{ .}\]
		\end{enumerate}
		The second case will follow analogously to the first one. 
		So, for the first case, 
		suppose 
		\begin{align}\label{thmC-randomMAGscase1}
		K\left( \mathscr{E}( \mathscr{G}_c ) \, \big\mid \left| \mathbb{V}( \mathscr{G}_c ) \right| \right) 
		\leq
		K\left( E( G ) \, \big\mid \left| V( G ) \right| \right)  + \mathbf{O}(1) 
		\text{ .}
		\end{align}
		Note that
		\begin{align*}
		& l\left(   \left< E( G ) \right>  \right)
		\leq
		\mathbf{O}\left( \left( \frac{ { \left| V( G ) \right| }^2 - { \left| V( G ) \right| } }{ 2 } \right)^2 \right)
		\end{align*}
		\noindent Therefore, from Lemma~\ref{lemmaBasicAIT}, we will have by supposition (see Equation~\eqref{thmC-randomMAGscase1}) that  
		\begin{align*}
		C\left( \mathscr{E}( \mathscr{G}_c ) \, \big\mid \left| \mathbb{V}( \mathscr{G}_c ) \right|\right) 
		& \leq
		K\left( \mathscr{E}( \mathscr{G}_c ) \, \big\mid \left| \mathbb{V}( \mathscr{G}_c ) \right| \right) 
		+ 
		\mathbf{O}(1)
		\leq 
		K\left( E( G ) \, \big\mid \left| V( G ) \right| \right)  + \mathbf{O}(1) 
		\leq \\
		& \leq
		C\left( E( G ) \, \big\mid \left| V( G ) \right| \right)
		+
		\mathbf{O}( \lg( \left| V( G ) \right| ) )
		\text{ .}
		\end{align*}
		\noindent For the second case
		\[
		K\left( \mathscr{E}( \mathscr{G}_c ) \, \big\mid \left| \mathbb{V}( \mathscr{G}_c ) \right| \right) + \mathbf{O}(1)
		\geq
		K\left( E( G ) \, \big\mid \left| V( G ) \right| \right)  
		\text{ ,}
		\]
		\noindent we will analogously have that 
		\[
		C\left( E( G ) \, \big\mid \left| V( G ) \right| \right)
		\leq
		C\left( \mathscr{E}( \mathscr{G}_c ) \, \big\mid \left| \mathbb{V}( \mathscr{G}_c ) \right| \right) 
		+
		\mathbf{O}( \lg( \left| \mathbb{V}( \mathscr{G}_c ) \right|) )
		\text{ .}
		\]
		\noindent To this end, just note that one can use the recursive function $ f^{-1} $ to construct the composite vertices in $ \mathbb{V}( \mathscr{G}_c ) $ from vertices in $ V( G )  $.
		Let $ F_G $ be a family of the classical graphs $G$ that are isomorphic to each MAG in $ F_{ \mathscr{G}_c } $.
		Thus, from Definition~\ref{defC-randomMAG} (and~\ref{defC-randomgraph}) and function $f$ applied to the recursively labeling of family $ F_{ \mathscr{G}_c } $, we have that 
		\begin{center}
			$ \mathscr{G}_c$ is $ \left( \delta( \left| \mathbb{V}( \mathscr{G}_c ) \right| ) + \mathbf{O}( \lg( \left| \mathbb{V}( \mathscr{G}_c ) \right|) ) \right) $-C-random \\
			\textit{iff} \\
			$ G $ is $ \left( \delta( \left| V( G ) \right| ) + \mathbf{O}( \lg( \left| V( G ) \right| ) ) \right) $-C-random, 
		\end{center}
		\noindent with respect to $ F_{ \mathscr{G}_c } $ and $ F_G $ respectively,
		 where $ \left| V( G ) \right| = \left| \mathbb{V}( \mathscr{G}_c ) \right|  \in \mathbb{N} $.
	\end{proof}
\end{thmafterlemmaundersection}

In addition, from Equations~\eqref{equationWeaklyKrandom} and~\eqref{equationStronglyKrandom} in Theorem~\ref{thmC-randomMAGsandgraphs} and from Definitions~\ref{defK-randomMAGs} and~\ref{defStronglyK-randomMAGs}, we will directly have that:

\begin{thmafterlemmaundersection}\label{BthmKrandomMAGsandgraphs}
	Let $  F_{ \mathscr{G}_c } \neq \emptyset  $ be an arbitrary recursively labeled family of simple MAGs $ \mathscr{G}_c $ (as in Definition~\ref{defLabeledfamilyofMAG}).
	Then, for every $  \mathscr{G}_c  \in  F_{ \mathscr{G}_c } $,
	\begin{center}
		$ \mathscr{G}_c$ is weakly/strongly $ \left( \delta( \left| \mathbb{V}( \mathscr{G}_c ) \right| ) + \mathbf{O}( 1 ) \right) $-K-random \\
		$ \text{iff} $ \\
		$ G $ is weakly/strongly $ \left( \delta( \left| V( G ) \right| ) + \mathbf{O}( 1 ) \right) $-K-random 
	\end{center}
	\noindent where $ G $ is isomorphic (as in Corollary~\ref{corClassicMAGisomorphism}) to $ \mathscr{G}_c $.
\end{thmafterlemmaundersection}

\section{Some topological properties of algorithmically random finite multiaspect graphs}\label{sectionTopologicalproperties}

In this section, we extend the results on classical graphs in \cite{Li1997,Buhrman1999} to plain algorithmically random MAGs. 
We will investigate diameter, connectivity, degree, and automorphisms.
To this end, Theorem~\ref{thmC-randomMAGsandgraphs} establishes a way to study common properties between algorithmically random MAGs and algorithmically random graphs. It takes into account algorithmic randomness for plain algorithmic complexity in both cases. In fact, we have shown that the plain algorithmic complexity of simple MAGs and its isomorphic classical graph is roughly the same, except for the amount of algorithmic information necessary\footnote{ Upper bounded by $ \mathbf{O}( \lg( \left| V( G ) \right| ) ) $.} to encode the length of the program that performs this isomorphism on an arbitrary universal Turing machine. As a consequence, it allows us to properly extend some results in \cite{Li1997,Buhrman1999} on plain algorithmically random classical graphs to simple MAGs:

\begin{corollaryafterlemmaundersection}\label{corC-randomMAGs}
	Let $  F_{ \mathscr{G}_c }  $ be an arbitrary recursively labeled infinite family of simple MAGs $ \mathscr{G}_c $ (as in Definition~\ref{defLabeledfamilyofMAG}).
	Then, the following hold for large enough $ \mathscr{G}_c \in F_{ \mathscr{G}_c } $:
	\begin{enumerate}
		\item If family $  F_{ \mathscr{G}_c } $  contains all possible arrangements of presence or absence of composite edges for any set of composite vertices $ \left| \mathbb{V}( \mathscr{G}_c ) \right| $ (e.g., the family in the proof of Lemma~\ref{lemmaLabeledfamilyofMAG}), then a fraction of at least 
		\[
		1 - \frac{1}{2^{ \delta( \left| \mathbb{V}( \mathscr{G}_c ) \right| )   }} 
		\]
		\noindent of all MAGs that belong to this family $  F_{ \mathscr{G}_c } $ is $ \delta( \left| \mathbb{V}( \mathscr{G}_c ) \right| + \mathbf{O}( \lg( \left| \mathbb{V}( \mathscr{G}_c ) \right|) ) ) $-C-random. \label{corC-randomMAGs1}
		
		\item The degree $ \mathbf{d}( \mathbf{v} ) $ of a composite vertex $ \mathbf{v} \in \mathbb{V}( \mathscr{G}_c ) $ in a $ \delta( \left| \mathbb{V}( \mathscr{G}_c ) \right|  ) $-C-random MAG $  \mathscr{G}_c  $ satisfies
		\[
		\left| \mathbf{d}( \mathbf{v} ) - \left( \frac{ \left| \mathbb{V}( \mathscr{G}_c ) \right| - 1 }{ 2 } \right) \right| 
		= 
		\mathbf{O}\left( \sqrt{ \left| \mathbb{V}( \mathscr{G}_c ) \right| \, \left( \delta( \left| \mathbb{V}( \mathscr{G}_c ) \right| ) + \mathbf{O}( \lg( \left| \mathbb{V}( \mathscr{G}_c ) \right|) )  \right) } \right)
		\]
		\label{corC-randomMAGs2}
		
		\item $ \mathbf{o}( \left| \mathbb{V}( \mathscr{G}_c ) \right|  ) $-C-random MAGs $  \mathscr{G}_c  $ have 
		\[ 
		\frac{\left| \mathbb{V}( \mathscr{G}_c ) \right| }{4} \pm \mathbf{o}(\left| \mathbb{V}( \mathscr{G}_c ) \right| )
		\]
		\noindent disjoint paths of length 2 between each pair of composite vertices $ \mathbf{u} , \mathbf{v} \in \mathbb{V}( \mathscr{G}_c ) $. In particular, $ \mathbf{o}( \left| \mathbb{V}( \mathscr{G}_c ) \right|  ) $-C-random MAGs $  \mathscr{G}_c  $ have composite diameter $2$.
		\label{corC-randomMAGs3}
		
		\item 
		Let $  \mathscr{G}_c  $ be $ ( \mathbf{O}\left( \lg(\left| \mathbb{V}( \mathscr{G}_c ) \right| ) \right) )$-C-random.
		Let $ X_{ f( \left| \mathbb{V}( \mathscr{G}_c ) \right|  ) }( \mathbf{v} ) $ denote the set of the least $ f(\left| \mathbb{V}( \mathscr{G}_c ) \right| ) $ neighbors of a composite vertex $ \mathbf{v} \in \mathbb{V}( \mathscr{G}_c ) $, where
		\[
		\myfunc{ f }{ \mathbb{N}}{ \mathbb{N} }{ \left| \mathbb{V}( \mathscr{G}_c ) \right|  }{ f(\left| \mathbb{V}( \mathscr{G}_c ) \right| ) }
		\]
		Then, for every composite vertices $ \mathbf{u} , \mathbf{v} \in \mathbb{V}( \mathscr{G}_c ) $,
		\begin{align*}
		\{ \mathbf{u} , \mathbf{v} \} & \in \mathscr{E}( \mathscr{G}_c ) \\
		& \text{or} \\
		\exists \mathbf{i} \in \mathbb{V}( \mathscr{G}_c ) \big( \mathbf{i}  \in  X_{ \left( \lg\left( \left| \mathbb{V}( \mathscr{G}_c ) \right| \right)  \right)^2  }( \mathbf{v} ) \land \{ \mathbf{u} , \mathbf{i} \} & \in \mathscr{E}( \mathscr{G}_c ) \land \{ \mathbf{i} , \mathbf{v} \} \in \mathscr{E}( \mathscr{G}_c ) \big) 
		\end{align*}
		\label{corC-randomMAGs4}
		
		\item $ \mathbf{o}( \left| \mathbb{V}( \mathscr{G}_c ) \right| - \lg( \left| \mathbb{V}( \mathscr{G}_c ) \right| )  ) $-C-random MAGs $  \mathscr{G}_c  $ are \emph{rigid} under permutations of composite vertices.\footnote{ That is, under a permutation of composite vertices that entails a reordering of the terms of the characteristic string such that the ordering of composite edges remains unchanged.} 
		\label{corC-randomMAGs5}
	\end{enumerate}

	\begin{proof}
		The proofs of all five statements come directly from Theorem~\ref{thmC-randomMAGsandgraphs}. Hence, we specifically obtain the desired proofs of Items~\ref{corC-randomMAGs1}, \ref{corC-randomMAGs2}, \ref{corC-randomMAGs3}, \ref{corC-randomMAGs4}, and \ref{corC-randomMAGs5} from Lemmas~\ref{lemmaFractionofrandomgraphs}, \ref{lemmaDegreerandomgraph},  \ref{lemmaDiameterrandomgraph}, \ref{lemmaStarinrandomgraphs},  and~\ref{lemmaRigidgraphs} respectively. Note that one needs to apply the respective corrections to the randomness deficiencies $ \delta(x) $ from Theorem~\ref{thmC-randomMAGsandgraphs}. Also note that, in Item~\ref{corC-randomMAGs4}, if a classical graph is $ \left(c \, \lg( \left| V( G ) \right| )\right) $-C-random, then $ \left( ( c + 3) \, \lg( \left| V( G ) \right| ) \right) \leq \mathbf{o}( \left( \lg( \left| V( G ) \right| ) \right)^2 ) $, which satisfies Lemma~\ref{lemmaStarinrandomgraphs}.
	\end{proof}
	
\end{corollaryafterlemmaundersection}

In addition, we can directly\footnote{ Hence, we omit the proof.} combine Corollary~\ref{corC-randomMAGs} with 
Corollary~\ref{thmK-randomandC-randomnestedsubMAGs} into respectively the following 
Corollary~\ref{corK-randomnestedfamilyproperties}, which can be easily extended to classical graphs too. 
These results end our present investigation of algorithmic randomness of multiaspect graphs (MAGs) by relating network topological properties with prefix algorithmically random MAGs.
For example, the results and theoretic methods developed in the present article also enable the mathematical investigation of other network topological properties that can only happen in the multidimensional case, as in \cite{Abrahao2019bnat}.

\begin{corollaryafterlemmaundersection}\label{corK-randomnestedfamilyproperties}
	Let $  F_{ \mathscr{G}_c }^{v*}   $ be a $ \mathbf{O}(1) $-K-random infinite (vertex-induced) nesting family (as in Theorem~\ref{thmNestedfamilyofK-randomMAGs}) of simple MAGs $ \mathscr{G}_c $. Then, the following hold for large enough $ \mathscr{G}_c \in F_{ \mathscr{G}_c }^{v*} $:
	\begin{enumerate}
		\item The degree $ \mathbf{d}( \mathbf{v} ) $ of a composite vertex $ \mathbf{v} \in \mathbb{V}( \mathscr{G}_c ) $ in a MAG $  \mathscr{G}_c $ satisfies
		\[
		\left| \mathbf{d}( \mathbf{v} ) - \left( \frac{ \left| \mathbb{V}( \mathscr{G}_c ) \right| - 1 }{ 2 } \right) \right| 
		= 
		\mathbf{O}\left( \sqrt{ \left| \mathbb{V}( \mathscr{G}_c ) \right| \, \left(  \mathbf{O}( \lg( \left| \mathbb{V}( \mathscr{G}_c ) \right|) )  \right) } \right)
		\]
		\label{corK-randomnestedfamilyproperties1}
		
		\item MAG $  \mathscr{G}_c $ has
		\[ 
		\frac{\left| \mathbb{V}( \mathscr{G}_c ) \right| }{4} \pm \mathbf{o}(\left| \mathbb{V}( \mathscr{G}_c ) \right| )
		\]
		\noindent disjoint paths of length 2 between each pair of composite vertices $ \mathbf{u} , \mathbf{v} \in \mathbb{V}( \mathscr{G}_c ) $. 
		\label{corK-randomnestedfamilyproperties2}
		
		\item MAG $  \mathscr{G}_c $ has composite diameter $2$.
		\label{corK-randomnestedfamilyproperties3}
		
		\item For every composite vertices $ \mathbf{u} , \mathbf{v} \in \mathbb{V}( \mathscr{G}_c ) $ in a MAG $  \mathscr{G}_c $,
		\begin{align*}
		\{ \mathbf{u} , \mathbf{v} \} & \in \mathscr{E}( \mathscr{G}_c ) \\
		& \text{or} \\
		\exists \mathbf{i} \in \mathbb{V}( \mathscr{G}_c ) \big( \mathbf{i}  \in  X_{ \left( \lg\left( \left| \mathbb{V}( \mathscr{G}_c ) \right| \right)  \right)^2  }( \mathbf{v} ) \land \{ \mathbf{u} , \mathbf{i} \} & \in \mathscr{E}( \mathscr{G}_c ) \land \{ \mathbf{i} , \mathbf{v} \} \in \mathscr{E}( \mathscr{G}_c ) \big) 
		\end{align*}
		\label{corK-randomnestedfamilyproperties4}
		
		\item MAG $  \mathscr{G}_c $ is rigid under permutations of composite vertices. 
		\label{corK-randomnestedfamilyproperties5}
	\end{enumerate}
\end{corollaryafterlemmaundersection}

\section{Conclusions}\label{sectionConclusion}
This article has presented mathematical results on the conditions, limitations, and concepts for algorithmic information theory to be applied to the study of multidimensional networks, such as dynamic networks or multilayer networks, in the form of multiaspect graphs (MAGs).

We have extended the conception of encodings in order to define and construct recursively labeled families of MAGs that do not display the algorithmic complexity distortions in \cite{Abrahao2020cpublished,Abrahao2021publishednat}.
In this case, we have shown that the algorithmic information of MAGs in a recursively labeled family are indeed tightly associated (analogously to the case for classical graphs) with its respective characteristic string.
This way, we have demonstrated that, although it does not hold in the general case (as described in the previous paragraph), there are particular infinite families of MAGs that are equivalent to their characteristic strings in terms of algorithmic information.
This formally grounds and enables the network compressibility/incompressibility analysis of several particular types of multidimensional networks in the same manner as the classical graph case.

We also introduced prefix algorithmic randomness for MAGs. 
Recursively labeled infinite nesting families of MAGs were formally constructed with the purpose of investigating an infinite family of finite multidimensional object that behaves like an infinite binary sequence.
Indeed, we have shown that, even in the multidimensional case, there is such an infinite nesting family in which each (vertex-induced) subMAG represented in a unified string-based encoding is incompressible with respect to prefix algorithmic complexity.


Furthermore, we have investigated the algorithmic-informational cost of the isomorphism between a MAG and its respective isomorphic graph.
Indeed, regarding the connection through composite edges, we have shown that not only ``most'' of the network topological properties of such graph are inherited by the MAG (and vice-versa), but also ``most'' of those that derives from the graph's topological incompressibility.  
These results set sufficient conditions for extending previous work on lossless incompressibility of graphs to the investigation of lossless incompressibility of MAGs,
so that one can achieve mathematical proofs of the existence of certain network topological properties in multidimensional networks, which are not necessarily generated or defined by stochastic processes.
As it was the case for classical graphs, this shows that there are several useful properties that could be embedded or analyzed in multidimensional networks.
In this sense, we suggest for future research the study of more inherited topological properties for such multidimensional networks, and possibly under other randomness deficiencies. 
Another interesting future topic of investigation would be estimating the number of possible topological properties that are inherited given a certain randomness deficiency. 

Discussions about, either in favor or against, the use of Shannon’s information theory with the purpose of investigating mathematical properties of graphs or complex networks cannot make further progress unless it is replaced or complemented by measures of algorithmic randomness. While it is true that physics, and many other areas, have been slow at moving away from and beyond Shannon, entropy can still find a wide range of applications but it forces researchers to keep tweaking and proposing a plethora of ad hoc measures of information to render their features of interest visible to the scope of their quantitative measures. Without the algorithmic component, however, they eventually fail at producing generative models from first principles, once they cannot distinguish between what can be generated recursively from what it cannot. In this way, algorithmic complexity and algorithmic randomness as salient properties in the interface between discrete mathematics, logic, and theoretical computer science have been showing to be fundamental not only to theoretical purposes, but also in the scientific method (especially, in the challenge of causality discovery) for studying complex networks. Fears against moving away from classical information based upon arguments of the (semi-)uncomputability of AIT should not preclude progress. For this reason, we also think this paper makes an important contribution to the study of computably irreducible information content or lossless incompressibility in multidimensional object representations.


\section{Financial support}

F. S. Abrahão acknowledges the partial support from São Paulo Research Foundation (FAPESP), grant \#$2021$/$14501$-$8$.
Authors also acknowledge the partial support from CNPq through their individual grants: F. S. Abrahão (313.043/2016-7), K. Wehmuth (312599/2016-1), and A. Ziviani (308.729/2015-3). Authors acknowledge the INCT in Data Science – INCT-CiD (CNPq 465.560/2014-8). Authors also acknowledge the partial support from CAPES/STIC-AmSud (18-STIC-07), FAPESP (2015/24493-1), and FAPERJ (E-26/203.046/2017). H. Zenil acknowledges the Swedish Research Council (Vetenskapsr\r{a}det) for their support under grant No. 2015-05299.

\bibliographystyle{plainnat}
\bibliography{3-CompleteRefs-Felipe.bib}


\end{document}